# Dissecting functional degradation in NiTi shape-memory-alloys containing amorphous regions via atomistic simulations

Won-Seok Ko[a,*], Won Seok Choi[b], Guanglong Xu[c], Pyuck-Pa Choi[b,**], Yuji Ikeda[d], and Blazej Grabowski[d]

[a] School of Materials Science and Engineering, University of Ulsan, Ulsan 44610, Republic of Korea
[b] Department of Materials Science and Engineering, Korea Advanced Institute of Science and Technology, 34141 Daejeon, Republic of Korea
[c] Tech Institute for Advanced Materials & College of Materials Science and Engineering, Nanjing Tech University, 210009 Nanjing, China
[d] Institute of Materials Science, University of Stuttgart, Pfaffenwaldring 55, 70569 Stuttgart, Germany

## Abstract

Molecular dynamics simulations are performed to provide a detailed understanding of the functional degradation of nano-scaled NiTi shape memory alloys containing amorphous regions. The origin of the experimentally reported accumulation of plastic deformation and the anomalous sudden increase of the residual strain under cyclic mechanical loading are explained by detailed insights into the relevant atomistic processes. Our work reveals that the mechanical response of shape-memory-alloy pillars under cyclic compression is significantly influenced by the presence of an amorphous-like grain boundary or surface region. The main factor responsible for the observed degradation of superelasticity under cyclic loading is the accumulated plastic deformation and the resultant retained martensite originating from a synergetic contribution of the amorphous and crystalline shape-memory-alloy regions. We show that the reported sudden diminishment of the stress plateaus and of the hysteresis under cyclic loading is caused by the increased stability of the martensite phase due to the presence of the amorphous phase. Based on the identified mechanism responsible for the degradation, we validate reported methods of recovering the superelasticity and propose a new method to prohibit the synergetic contribution of the amorphous and crystalline regions, such as to achieve a sustainable operation of shape memory alloys at small scale.

**Keywords:** Shape memory alloy, Molecular dynamics, Phase transformation, Nanopillar, Nickel-Titanium

* Corresponding author: wonsko@ulsan.ac.kr (Won-Seok Ko)
**Corresponding author: p.choi@kaist.ac.kr (Pyuck-Pa Choi)

# 1. Introduction

Shape memory alloys (SMAs) are a category of materials with the property of recovering their initial shape upon heat treatment (shape memory effect) and of sustaining large elastic strains upon mechanical loading (superelasticity). In SMAs, the reversible temperature- and stress-induced martensitic phase transformation between austenite and martensite results in the shape memory effect and in superelasticity, respectively. These unique properties render SMAs eminently suitable for various functional applications [1-4]. Nevertheless, critical challenges in applying SMAs remain, in particular overcoming the degradation of performance under cyclic loading, which can severely limit the service life of SMA parts. In the bulk state, cyclic deformation of SMAs usually leads to functional degradation as characterized by a gradual increase in the residual strain, decrease in the stress required for the austenite to martensite transformation, and the resultant decrease in the hysteresis loop area [3].

Recently, applications of SMAs in micro- and nano-electromechanical systems (MEMS/NEMS) have actively been pursued [5, 6], e.g., as effective actuators, sensors, or mechanical damping materials. For these applications, methods for manufacturing SMA structures at the small scale using focused ion beam (FIB) technology are critical [7, 8]. As the properties of miniaturized SMAs can differ significantly from their bulk counterparts, their distinctive characteristics of deformation and phase transformation at the nano- and micron-scale are of great interest. Corresponding experimental studies have been therefore pursued [5, 6]. In particular, the size dependence of the stress-induced phase transformation and resultant superelasticity have been studied by nano-/micro-compression tests of single crystal pillars prepared by FIB [5, 9-16]. Moreover, the cyclic phase transformation behavior of nanocrystalline NiTi wires, pillars and tubes was investigated [17-21]. From the theoretical side, atomistic simulations such as molecular dynamics (MD) combined with semi-empirical interatomic potentials have been employed to provide a detailed understanding of the underlying mechanisms to supplement experiments. In particular, MD simulations were successfully applied to examine the distinctive features of NiTi SMAs at the nanoscale, e.g. for nanocrystalline SMAs [22-24], SMA nanoprecipitates embedded in a stiff non-transforming matrix [25], freestanding SMA nanoparticles [26], SMA nanowires [27], SMA nanopillars [28, 29], and SMAs with complex microstructures [30] based on a reliable interatomic potential [31, 32].

Unfortunately, the previous studies [9-12, 15] revealed that NiTi SMAs at the small scale suffer from a significant deterioration of superelasticity, even much stronger than their bulk counterparts. SMAs prepared at the small scale generally exhibit an incomplete recovery of the initial shape after mechanical loading and unloading and show a residual strain in their stress-strain response. Incomplete recovery is a problem of bulk SMAs as well [3], however, it is significantly more pronounced when the size of SMAs decreases. For example, based on a study of single crystal NiTi pillars with diameters between 149 nm and 1990 nm Frick *et al*. [9, 10] revealed appreciable residual strain after compression in the smaller sized pillars and a full loss of superelasticity for pillars smaller than 200 nm. Moreover, maintaining functional stability under cyclic loading seems to be a serious problem for small scale SMAs. For example, previous works on nanocrystalline NiTi wires by Delville *et al.* [17, 18], nanocrystalline NiTi pillars by Hua *et al.* [21], and nanocrystalline NiTi thin-walled tubes by Ghassemi-Armaki *et al.* [19] reported functional degradation under cyclic mechanical loading in a manner similar to bulk NiTi SMAs (i.e., a gradual increase in the residual strain and decrease in the transformation stress and hysteresis loop area).

One possible explanation for the degradation of superelasticity of small-scale SMAs is the occurrence of plastic deformation associated with dislocation slip. For example, a certain fraction of dislocations could already be present in the SMA and could be activated during deformation, or new dislocations could be nucleated during mechanical loading. This argument is plausible for explaining the occurrence of the residual strain as dislocation slip can cause irreversible macroscopic deformation, which needs to be distinguished from superelastic behavior. In fact, previous research explained that the functional degradation of bulk SMAs is governed by the lattice-level strain incompatibility between the austenite and martensite phase [2]. Such a strain incompatibility leads to localized, high stresses and to the formation of dislocations at the austenite-martensite interfaces [12, 33, 34]. For the single crystal NiTi micropillars, this explanation is also plausible as previous work by Norfleet *et al.* [12] revealed that transformation-induced dislocations can cause functional degradation of NiTi micropillars subjected to just a few loading cycles. Several experimental investigations of NiTi SMAs at the small scale [11-14, 21] demonstrated the presence of dislocations. A previous MD simulation on NiTi nanopillars by Wang *et al.* [29] also revealed that dislocations observed during cyclic deformation are responsible for the degradation of the superelasticity.

However, there are other experimental results that cannot be explained solely by the effect of dislocations. According to the already above-mentioned work of Frick *et al.* [10], the residual strain of smaller pillars is greater than that of larger pillars regardless of the loading orientation. Based on the explanation that the dislocation activity is a main source of the residual strain, a more pronounced residual strain of the smaller pillars can be ascribed to an enhanced dislocation activity. However, this explanation is inconsistent with the trend of 'smaller is stronger', which is well-known in the field of small-scale deformation. It is known that the dislocation activity is suppressed with decreasing pillar size due to dislocation starvation [35, 36] and limited size of dislocation sources [11, 37, 38]. Moreover, if the dislocation activity was a main source of the residual strain, a dependency on the loading orientation should be present because of different effective stresses for the activation of dislocations. However, experimental studies [10] have not reported any orientation dependency.

Another possible factor for the degradation of superelasticity in SMAs at the small scale is the occurrence of plastic deformation induced in or at regions with an amorphous character. For example, as general high-angle grain boundaries locally exhibit structural disorder, nanocrystalline SMAs with a high grain boundary to volume ratio can be classified as composite materials of crystalline and amorphous regions. It is well known that the governing plastic deformation of nanocrystalline materials can change from dislocation slip to grain boundary sliding with decreasing grain sizes. Indeed, a previous MD study on nanocrystalline NiTi [22] confirmed that localized plastic deformation at grain boundaries is related to the residual strain after a loading and unloading cycle. Moreover, it is well known that high energy ion beams used for FIB milling can induce the formation of an amorphous surface layer. Several previous experimental investigations of NiTi SMAs also demonstrated the presence of an amorphous surface layer after FIB milling [9-11, 13, 39]. Because the thickness of the FIB-induced amorphous region is nearly independent of the pillar size, this amorphous region increasingly contributes to the deformation as the size of the pillar decreases. If indeed plastic deformation occurred in this amorphous region, a much more severe degradation of superelasticity for pillars with smaller dimensions and smaller grain sizes would be the consequence, because of the increased proportion of the amorphous region. This explanation seems convincing and consistent with the reported experimental trend for comparably small pillars at the nanoscale [9, 10]. In micron-sized pillars, however, the proportion of the grain boundary region and the FIB-induced amorphous layer (15 – 20 nm [9, 10]) seems to be too small to play a dominant role in the degradation of superelasticity.

The present study aims to provide a comprehensive understanding of the degradation of the superelasticity of SMAs at the small scale by analyzing atomistic details of the deformation and phase transformation behavior using MD simulations. In contrast to previous investigations focusing on the role of dislocation slip, we especially focus on the role of plastic deformation caused by amorphous regions and the resulting impact on the functional degradation of SMAs at the small scale, all of which cannot be readily analyzed by experiments. We mimic in particular the impact of grain boundaries and FIB milling in our simulations by considering nanocrystalline pillars and single crystal pillars with an amorphous surface shell in the initial simulation setup before the loading. Based on this preparation, we reproduce the experimentally reported accumulation of plastic deformation and the anomalous sudden increase of the residual strain under cyclic loading. We discuss the underlying mechanisms responsible for the observed degradation of superelasticity and suggest a possible solution for obtaining a reliable performance of SMAs at the small scale.

## 2. Methodology

For the present study, we have selected a nickel-titanium (NiTi) SMA with an equiatomic composition as the material for our pillars. In NiTi alloys, the reversible temperature- and stress-induced phase transformation between cubic B2 (austenite) and monoclinic B19' (martensite) results in the shape memory effect and in superelasticity, respectively [2].

The present MD simulations were conducted using the LAMMPS code [40] with a previously developed second nearest neighbor modified embedded-atom method (2NN MEAM) interatomic potential for the binary Ni-Ti system [31]. This interatomic potential was specifically designed to accurately reproduce the temperature- and stress-induced phase transformation between the B2 austenite and B19' martensite phase in the equiatomic NiTi alloy. The reliability of this interatomic potential has been verified by previous MD studies [22, 26, 41-43] focusing on phase transformations and microstructural evolution. A radial cutoff distance of 5.0 Å, which is larger than the second nearest-neighbor distance of B2 and B19' structures, was used for all simulations of the present study.

Initially, pristine single crystal pillars with a circular cross section and different lateral diameters (15, 20, 25 and 30 nm) were prepared using the crystal structure of the B2 austenite phase for the *whole* pillar. To analyze the influence of crystal orientation on superelasticity, different pillars were prepared based on selected crystallographic loading directions, specifically [001], [111], [112] and [011]. All the pillars had the same aspect ratio of 1.0 (height/width). Periodic boundary conditions were applied along the loading axis to minimize the influence of the aspect ratio on the predicted deformation behavior. Based on benchmark simulations using single crystal pillars with aspect ratios of 2.0 and 3.0, we found that the aspect ratio has only a marginal effect on the critical stress of the austenite to martensite phase transformation and the elastic deformation of pillars [Fig. 17(a) in the Appendix]. Schematic configurations of pillars used in the present study are shown in Fig. 1. The resulting cell dimensions and numbers of atoms for each pillar are summarized in Table 1.

To mimic experimental conditions of single crystal pillars fabricated by FIB milling, a surface shell with an amorphous structure was added on top of the pristine pillar with [001] loading direction (see Sec. 3.1 for the justification of this choice), based on the assumption that a FIB induced region does not have a clear crystalline structure. Different thicknesses (0.5, 1.0, 1.5, 2.0, 2.5 and 3.0 nm) of the amorphous surface shell were applied to pillars of different diameters (15, 20, 25 and 30 nm) as shown in Fig. 1. The amorphous region was prepared by randomly disturbing the original positions of the atoms in the crystalline structure with a maximum displacement of 0.2 nm.

A nanocrystalline pillar with a diameter of 30 nm was generated using the Voronoi construction method [44] with random positions and crystallographic orientations for each grain. Initially, a cube-shaped nanocrystalline cell with the B2 austenite structure was generated considering a designated number of grains ($N$: 5) and average diameter ($D$: 22 nm). A nanocrystalline pillar with a circular cross section was then prepared by cutting the cube-shaped cell as shown in Fig. 1.

The generated pillars were subjected to an energy minimization process using the conjugate gradient method to obtain equilibrium configurations. A series of MD simulations was then performed with a time step of 2 fs in an isobaric-isothermal (*NPT*) ensemble at the designated temperature and pressure (stress) along the loading direction. The Nosé-Hoover thermostat and barostat [45, 46] were used for controlling temperature and pressure, respectively. A relaxation at a temperature above the austenite finish ($A_f$) temperature was performed to maintain the initial B2 austenite structure. A stress-controlled uniaxial loading was then applied by adjusting the stress along the longitudinal direction of the pillar. The compressive stress was gradually increased to the maximum value and decreased to zero, and the cell dimensions and individual atomic positions were allowed to fully relax under a given stress state. Note that due to the thermostatting during the deformation and transformation, the present MD simulations correspond to isothermal boundary conditions, whereas typical experimental conditions are intermediate between isothermal and adiabatic due to the latent heat of the phase transformation. The effect of different thermal boundary conditions on the cyclic deformation of NiTi SMAs is well documented by previous MD works of Wang et al. [24, 47]. According to these works, the overall stress values obtained under adiabatic conditions are generally higher than those obtained under isothermal conditions. This is because the temperature increases with increasing number of cycles due to the accumulated latent heat under adiabatic conditions. To analyze the effect of the loading and unloading rates, benchmark simulations were performed for single crystal pillars loaded along the [001] direction. The resultant stress-strain response [Fig. 17(b) in the Appendix] indicates that the critical stress for the austenite to martensite transformation and the overall stress level are well converged for stress rates below 3.18 MPa/ps. Considering this convergence behavior and the computational requirements, this value was selected and applied for all simulations.

The microstructural evolution was determined using the adaptive cutoff common-neighbor analysis (AC-CNA) algorithm [48] and Polyhedral Template Matching (PTM) method [49] as implemented in the OVITO program [50]. Although these algorithms were initially not developed to differentiate between the B2 austenite and B19' martensite phases, we confirmed that they are well suited to identify the occurrence of the phase transformation. In the AC-CNA and PTM patterns, atoms depicted in blue represent the B2 austenite structure. The B19' martensite structure is mostly represented by red atoms. Undetermined regions, i.e., surfaces and amorphous regions, are shown in gray. Some atoms in the crystalline regions are also represented by gray color due to the incompleteness of the visualization method and the thermal noise at finite temperatures. The von-Mises local shear invariant [51] of each atom was calculated and visualized using the OVITO program to emphasize the local plastic deformation.

## 3. Results and discussion

### 3.1. Stress-induced phase transformation of pristine SMA pillars

To have a reference for investigating the impact of the amorphous region, it is helpful to analyze the deformation and phase transformation behavior of pristine, i.e., defect free, single crystal SMA pillars. We thus first examine the deformation and transformation characteristics of such pristine single crystal pillars under several selected loading orientations ([001], [011], [111] and [112]). Figure 2(a) shows the stress-strain responses of uniaxially compressed pillars with these orientations and a diameter of 20 nm obtained by MD simulations at 450 K. All stress-strain responses indicate a typical sigmoidal (i.e., S-shaped) stress-strain behavior typical for SMAs, reflecting the occurrence of a stress-induced phase transformation. Since the pillars were subjected to the compressive loading at a temperature higher than the martensite start ($M_s$) temperature, the thermodynamically stable B2 austenite transformed to the B19' martensite phase which explains the plateau in the curve. During unloading, the B19' martensite phase transformed back into the B2 austenite phase showing another plateau at a lower stress level. The resultant stress-strain curves indicate a complete hysteresis loop without any noticeable residual strain.

Figure 2(b) shows representative atomic configurations of pillars at different stress levels, i.e., the initial state, the state of maximum compressive loading, and the stress-free state after unloading. At the initial stage of loading, all pillars exhibit clearly the austenite structure (blue colored atoms). At maximum compressive loading, the austenite phase has mostly transformed into the martensite phase (red colored atoms) except for the [011] loading. In some cases, thin regions of the austenite phase remain in between the martensite phase at maximum loading [marked in Fig. 2(b) by "Domain boundaries"]. These regions correspond to martensitic domain boundaries which divide multiple martensite variants (domains). The microstructure of the martensite phase obtained after loading is shown in more detail in Fig. 2(c). Although the initial austenite phase is a single crystal, the transformed martensite phase is composed of a twinned B19' structure with finely dispersed (001) compound twin boundaries. In some cases, this twinned structure occurs with multiple variants (domains) divided by domain boundaries. When the nucleation of martensite begins at multiple sites of the pillar simultaneously, there should be an orientation mismatch between regions originating from different martensite nuclei. If this orientation mismatch is not resolved during the process of the martensite growth, a residual austenite (or domain boundary) can be present in the pillar. Based on several additional MD runs, we have confirmed that the occurrence and morphology of multiple variants is of a stochastic nature. A similar, twinned microstructure was also observed in previous experiments on the phase transformation of NiTi SMAs at the nanoscale [52]. It is interesting to note that the present MD simulations of the stress-induced transformation reveal the formation of a twinned martensite structure almost over the entire region of the pillar. In bulk SMAs, a twinned martensite structure is generally expected to form during a temperature-induced transformation, while during a stress-induced transformation, detwinned or reoriented martensite with a specific variant is formed depending on the loading direction [3]. The present results imply that the process of martensite twinning/detwinning with the distinctive twinned structure at the small scale can differ significantly from its bulk counterparts during a stress-induced transformation.

As shown in Fig. 2(b), the [011] loading is different from the other loadings as it results in the transformation between the B2 (blue atoms) and $L1_0$ structure (light green atoms). The presence of the $L1_0$ phase could be an artifact of the interatomic potential. Based on separate density functional theory calculations, we confirmed that the $L1_0$ structure is unstable at 0 K. In contrast, the calculation based on the interatomic potential predicts the $L1_0$ structure to be metastable with a slightly higher (0.15 meV/atom) structural energy than the B2 structure.

Overall, the stress-strain responses shown in Fig. 2 indicate that the phase transformation of single crystal pillars is strongly anisotropic with respect to different compressive loadings. The differences can be explained by the presence of preferentially and non-preferentially oriented martensite variants with respect to the elastic deformation and stress-induced phase transformation. If we compare the stress required for the transformation during compressive loading [Fig. 2(a)], [001] is the most preferred compression direction for the phase transformation. In the remaining part of the present work, we therefore focus on the [001] loading of pillars because a lower level of the transformation stress and a therefore reduced maximum loading is beneficial to reduce the computational burden. Moreover, the selection of this loading orientation for practical applications should be advantageous to obtain an optimal superelasticity of SMAs by avoiding possible plastic deformation due to lower transformation stresses.

By focusing on the [001] loading direction, we have evaluated the effect of temperature on the deformation and phase transformation behavior of pristine single crystal pillars. Figure 3 shows the stress-strain response of the pillar with a diameter of 15 nm at different temperatures (450 – 650 K). The forward and backward transformations are characterized by the upper and lower plateaus in the curve, respectively. Detailed processes of the transformation under loading and unloading are revealed by the atomic configurations at different stress levels and temperatures to the right of Fig. 3. The level of the stress of the plateau increases with increasing temperature, and the phase transformation is completely inhibited at the maximum investigated stress of 1.27 GPa if the temperature is very high (650 K). Each stress-strain response draws a nearly closed curve without any significant residual strain, regardless of the temperature.

### 3.2. Deformation and transformation behavior of SMA pillars containing amorphous regions

In the previous section, the transformation behavior of pristine single crystal pillars was examined, and the presence of an upper and lower plateau in the stress-strain response was revealed. Furthermore, no residual strain was observed. In this section, we extend the investigation to pillars containing amorphous regions and show that the deformation and transformation behavior is considerably different. We analyze two cases of pillars with amorphous structures: (i) a single crystal pillar with a surface amorphous shell (loaded in the [001] direction) and (ii) a nanocrystalline pillar with grain boundary regions.

Figure 4 shows the stress-strain response of the single crystal pillar with a surface amorphous shell at different temperatures (450 – 650 K). The pillar consists in the beginning of the loading stage of an austenitic single crystal core and an amorphous surface shell with a thickness of 0.5 nm (identified as gray atoms in Fig. 4). For a better-defined comparison with the pristine pillar, the surface shell was added on top of the original pristine pillar (15 nm) along the radial direction. The other simulation parameters (i.e., dimensions of the crystalline region, temperature, and maximum load) were kept the same as those in the previous section for the pristine pillar (Fig. 3). Figure 5 showcases the stress-strain response of the nanocrystalline pillar. A larger diameter of 30 nm was used to avoid too small grain sizes. Nevertheless, as shown in the cross section of the atomic configurations, the pillar contains a significant fraction of the amorphous grain boundary region (identified as gray atoms in Fig. 5). For the stress-strain response of the nanocrystalline pillar, a different temperature range was chosen (350 – 550 K, i.e., 100 K lower than for the single crystal pillars, 450 – 650 K), because we observed a substantial decrease in transformation temperatures [see Fig. 18(a,b) vs. (d) in the Appendix]. Decreased transformation temperatures for nanocrystalline SMAs were also reported by previous experiments [6, 53, 54].

Figure 4 reveals several characteristic features of the single crystal pillar with the amorphous shell. First of all, we observe an increase in the stress required for the forward transformation and also a reduction of the length of the corresponding plateau. This result reveals that it is more difficult to initiate the forward transformation in the pillar with the amorphous shell than in the respective pristine pillar. One possible explanation of the increased transformation stress could be an increase of the elastic modulus (i.e., the initial slope of the stress-strain curve) due to the presence of the amorphous shell. However, our calculations show only a marginal difference between the modulus of the pristine pillar (46.0 GPa) and the pillar with the amorphous shell (45.8 GPa). A more convincing explanation is that the amorphous region acts as a mechanical constraint and suppresses the forward transformation. Previous experimental [55] and simulation [22] works support this explanation. For the nanocrystalline pillar (Fig. 5), although a direct comparison to the pristine pillar is not possible due to differences in the diameter and temperature, likewise an increase in the stress required for the forward transformation and the disappearance of the plateau are clearly visible. Indeed, previous experimental [55] and simulation [22] works already revealed the impact of mechanical constraints on transformation stresses and the disappearance (or a reduction of the length) of the plateau in the stress-strain response.

Another significant feature of the pillars containing the amorphous regions as revealed by Figs. 4 and 5 is the presence of residual strain (remnant strain after unloading, i.e., at zero external stress), whereas the pristine pillar showed almost full recovery after unloading (cf. Fig. 3). This feature is consistent with reported experimental trends at the nanoscale [9, 10]. Because our simulations were performed without any initial dislocations and nucleation of new dislocations was not detected, the observed residual strain must be a direct consequence of plastic deformation of the amorphous regions. The deformation results at the highest temperatures (i.e., 650 K for the single crystal pillar with the amorphous shell and 550 K for the nanocrystalline pillar, Figs. 4 and 5 respectively) support this explanation. At these highest temperatures, neither a phase transformation nor nucleation of dislocations is observed, yet residual strain still occurs and can only be interpreted by the plastic deformation of the amorphous shell region. We performed additional simulations using single crystal pillars with different diameters (15, 20, 25, and 30 nm) and fixed thickness of the amorphous shell (0.5 nm) to investigate the size effect (i.e., greater residual strain in smaller pillars) observed in previous experiments [9, 10]. The results shown in Fig. 6(a) indeed confirm the validity of the hypothesis that the residual strain is more pronounced in smaller pillars because the FIB-induced amorphous shell region occupies a higher proportion of the pillar. We further substantiated the explanation that the plastic deformation is carried by the amorphous shell by investigating the accumulated atomic local shear strains after the unloading. As shown in Fig. 6(b), the plastic deformation is indeed localized at the surface shell region.

Figures 4 and 5 show that the amount of the residual strain is dependent on the temperature, i.e., the residual strain is less pronounced at higher temperatures. This result implies that the plastic deformation of the amorphous structure is less pronounced at high temperatures, which appears to conflict with the generally expected notion that materials soften with increasing temperature. This counterintuitive behavior can be explained by considering that the overall, maximally reachable strain *decreases* at a certain stress level with increasing temperature because the superelastic behavior is suppressed. This can be observed in Figs. 4 and 5 by inspecting for example the level of the strain at a fixed stress (e.g., 1.27 GPa) for the different temperatures. At low temperatures, the large strains induce large deformations of the amorphous region. This contribution decreases with temperature and thus counterbalances the increased temperature-induced plastic deformation of the amorphous region.

The deformation behavior at the lowest temperatures (450 K in Fig. 4 and 350 K in Fig. 5) reveals an exceptionally large residual strain. This result indicates that the plastic deformation of the amorphous shell is not the only factor determining the residual strain. In fact, we can clearly observe from Figs. 4 and 5 a fraction of retained martensite after unloading at this low temperature, contrary to the other temperatures. The presence of this retained martensite significantly contributes to the residual strain. Based on additional MD simulations on the temperature-induced transformation using an undeformed pillar, we confirmed that the presence of retained martensite in the nanocrystalline pillar at the lowest temperature (350 K in Fig. 5) can be readily explained by a higher $A_f$ temperature (370 K) of this pillar [see Fig. 18 (d) in the Appendix]. However, the presence of retained martensite of the single crystal pillar with the amorphous shell (450 K in Fig. 4) cannot be explained in this manner. We confirmed that the transformation temperatures of the pillar with the amorphous shell do not change as compared to the pristine pillar [see Fig. 18(a) vs. (b) in the Appendix]. In particular the $A_f$ temperature of the pillar with the amorphous shell is 420 K, i.e., below the temperature of 450 K at which the deformation has been performed (Fig. 4).

A possible explanation of the unexpected presence of the retained martensite is that the plastic deformation of the amorphous region serves to stabilize the martensite phase after unloading. This argument is clearly supported by comparing Figs. 3 and 4. For the reverse (backward) martensite to austenite transformation, the single crystal pillar with the amorphous shell exhibits decreased stress values and less pronounced plateaus as compared to the pristine single crystal pillar at the same temperature. Thus, the presence of the amorphous shell inhibits the backward transformation to austenite. As we already explained, the plastic deformation occurs in the amorphous region during the forward transformation. The deformed amorphous region appears to act as a constraint and to suppress the recovery of nearby crystalline SMA regions during the reverse transformation.

## 3.3 Deformation and transformation of SMA pillars containing amorphous regions during cyclic loading

In the previous section, we have shown that pillars with amorphous regions exhibit a deformation and phase transformation behavior distinct from their pristine counterparts under a *single* compressive loading and unloading process. The differences manifested themselves in the stress-strain response, in particular by the presence of residual strain after unloading and by altered transformation stresses. In this section, we extend our study to the deformation and phase transformation behavior of SMA pillars with amorphous regions under *cyclic* loading. All simulation conditions applied in this section correspond to those of the previous section except for the number of cycles ($n > 1$).

Figure 7 shows the stress-strain response of the [001] oriented single crystal pillar with an amorphous shell (same as the one in Fig. 4) for 5 cyclic loadings at different temperatures (400 – 650 K). The reference for the given strain values in each cycle is the longitudinal direction of the initial pillar. Corresponding atomic configurations in the initial stage and after each cycle are presented in Fig. 8. The results include simulations at two extreme temperature conditions: 400 K [Fig. 7(a)] and 650 K [Fig. 7(j)]. Since 400 K (a temperature not considered in the previous section for the pillar with the amorphous shell) is lower than the $A_f$ temperature (420 K) of the single crystal pillar, a loading at this temperature causes a spontaneous forward transformation and the reverse transformation is completely inhibited. Loading at 650 K does not involve any phase transformation of the pillar as already explained in the previous section. Figure 9 shows the stress-strain response of the nanocrystalline pillar for 5 cyclic loadings at different temperatures (300 – 550 K). Corresponding atomic configurations are presented in Fig. 10. The reverse transformation is completely inhibited during the loading at 300 K and 350 K because the $A_f$ temperature of this nanocrystalline pillar is 370 K. The loading at 550 K does not involve any phase transformation.

A notable common trend revealed by Figs. 7 and 9 is that the transformation stresses decrease significantly after the first cycle and continue to decrease as the number of cycles increases. Fig. 11 summarizes the key properties extracted from Fig. 7, i.e., from the stress-strain behavior of the single crystal pillar with the amorphous shell. Both forward and backward transformation stresses decrease as a function of the number of cycles [Figs. 11(a) and (b)], however the stress reduction during the forward transformation is significantly greater. Consequently, the hysteresis loop area, determined by the overall stress values during the forward and backward transformation, decreases with increasing number of cycles as illustrated in Fig. 11(e). A further feature of the cyclic loading simulations revealed by Figs. 7 and 9 is a steady increase of the residual strain with the number of cycles as exemplified in Fig. 11(c). Similarly, the maximum reachable strain, i.e., the maximum strain at the maximum stress (1.27 GPa), increases with each loading cycle [see in particular the magnified plots in Figs. 7(k) to (o), Figs. 9(k) to (o), and Fig. 11(d)].

Figure 11(f) shows a correlation plot between the residual strain and the maximum reachable strain obtained after the maximum (5$^{th}$) cycle for the case of the pillar with the amorphous shell. Two distinct regions corresponding to temperatures above and below about 480 – 490 K can be distinguished, each showing a close to linear correlation but with strongly differing slopes. At the higher temperatures, there is little increase of the residual strain with the maximum strain, whereas at the lower temperatures an extremely steep dependence is observed (note that temperatures increase from top right to bottom left). The occurrence of these two different correlation regimes at lower and higher temperatures is related to the presence or absence of retained martensite after cyclic loading.

The results of cyclic loading at higher temperatures [480 – 650 K in Fig. 7(e-j) and 400 – 550 K in Fig. 9(f-j)] do not exhibit any retained martensite, as explicitly confirmed by the atomic configurations in Figs. 8(e-j) and 10(f-j). The residual strain at these temperatures is thus a consequence of the plastic deformation of the amorphous shell and the amorphous grain boundary regions as discussed in the previous section. This deformation is enhanced when the martensitic transformation can occur during loading [480 – 600 K in Fig. 7(e-i) and 400 – 500 K in Fig. 9(f-i)] and the corresponding maximum strain shows a stronger dependence on the number of cycles than for the case where no transformation is possible (650 K in Fig. 7 and 550 K in Fig. 9). The stress-strain responses featuring the transformation [Figs. 7(e-i) and 9(f-i)] show the well-known training effect commonly observed in SMA systems. In previous experimental works on SMAs [3, 4], the training effect was in particular characterized by a stabilization of the hysteretic response and continuous decrease in the overall stress with cyclic loading, similarly as we observe for our simulated stress-strain responses [Figs. 7(e-i) and 9(f-i)].

In contrast, the cyclic loading at lower temperatures [400 – 470 K in Figs. 7(a-d) and 300 – 390 K in Figs. 9(a-e)] features retained martensite during cyclic loading, as can be seen from the atomic configurations in Figs. 8(a-d) and 10(a-e). Retained martensite can occur already after the first cycle or with a delay after several cycles, e.g., at 470 K for the single crystal pillar with the amorphous shell [Fig. 8(d)] and 390 K for the nanocrystalline pillar [Fig. 10(e)], and its fraction increases as the cyclic loading progresses. The loadings at temperatures below the $A_f$ temperature [400 K in Fig. 7(a) and 300 K and 350 K in Figs. 9(a,b)] are somewhat special and will be considered separately below. The other stress-strain responses exhibit comparably large residual strain values with a strong dependence on the number of cycles. Clearly, the retained martensite stored in the pillar from a preceding cycle plays a decisive role in changing the stress-strain response of subsequent cycles even if the fraction is small.

For example, as already mentioned, the loading of the single crystal pillar with the amorphous shell at 450 K [Fig. 4 in Sec. 3.2 and blue line in Fig. 7(b)] exhibits retained martensite already after a single loading cycle. In the second cycle [green lines Fig. 7(b)], the stress-strain response is impressively changed, even including a 'sudden' diminishment of both plateaus and thus of the enclosed hysteresis area. Further cycles influence the stress-strain curve less but the fraction of retained martensite gradually increases. A similar analysis can be performed for the stress-strain response of the single crystal pillar with the amorphous shell at 460 K, albeit with a delayed formation of retained martensite with respect to the number of cycles and hence with a delayed diminishment of the plateaus. At 470 K, the retained martensite sets in at an even later cycle and only the lower plateau related to the back transformation disappears after the fifth cycle. It is possible that further cyclic loading would modify also the plateau related to the forward transformation.

The deformation behavior under cyclic loading of the single crystal pillar with the amorphous shell at 480 K cannot be conclusively clarified based on the studied five loading cycles. A substantial increase in the number of cycles would be necessary to find out whether cyclic loading can lead to the formation of retained martensite and to a modification of the phase transformation plateaus for the 'intermediate' temperature of 480 K or whether the hysteresis loop stabilizes at a shape similar to the one observed after the fifth cycle. Corresponding simulations are beyond the scope of the present study.

The deformation behavior under cyclic loading at intermediate temperature ranges [450 – 470 K in Figs. 7(b-d) and 370 – 390 K in Figs. 9(c-e)] observed here can be utilized to explain the previous experimental results by Gómez-Cortés *et al.* [15] on CuAlNi SMA pillars. Specifically, this work [15] reported a loss of superelasticity quantified by a sudden increase of residual strain after a certain number of cycles (specifically after the 104[th] cycle) although the initial (first) cycle started with an almost complete recovery of the original shape. This anomalous sudden increase in residual strain was more pronounced when the number of cycles or the maximum load increased. The loss of superelasticity and the sudden increase of residual strain are consistent with the features extracted from the present cyclic loading simulations. The difference in the number of cycles (a few for the simulations and many more for the experiments) is related to the difference in pillar sizes accessible by experiment and simulation as will be discussed in more detail in the next section.

Due to limited information about the microstructure, the original experimental work [15] could only speculate that the origin of the anomalous residual strain is a sudden stabilization of the martensite phase due to dislocations created during cyclic loadings. A follow-up experimental work under nearly identical experiment conditions [16] did not exhibit any sudden increase in the residual strain. The reported experimental results thus signify that the overall deformation behavior of SMA pillars is very sensitive to the microstructure (e.g., surface conditions) and thus to the specific preparation process (e.g., FIB milling). The sensitivity of the transformation behavior of nano-scaled SMA pillars is nicely consistent with our simulation results, in particular with the results for the intermediate temperature range.

In our simulations at temperatures which lie below the $A_f$ temperature of the respective pillars [400 K in Fig. 7(a) and 300 K in Fig. 9(a)], we observe that the stress-strain response is nearly unchanged from the 2[nd] cycle on. Since the back transformation is completely suppressed already during the unloading in the first cycle, the crystalline part of the pillar remains in the martensite phase [Figs. 8(a) and 10(a)]. However, as shown in Fig. 2(c) and discussed previously, the transformed martensite phase is composed of a complex twinned structure. Due to the low energy required to switch between martensite variants, the martensitic microstructure can be reshaped during the ongoing cyclic loading in response to the plastic deformation of the amorphous shell. This explains the small variations in the maximum reachable and in the residual strain [Figs. 7(k) and 9(k)].

Overall, our cyclic loading simulations reveal an intriguing synergetic interplay of the crystalline and amorphous pillar parts during cyclic loading. As shown in Figs. 7(o) and 9(o), the stress-controlled cyclic loading at high temperatures (650 K in Fig. 7 and 550 K in Fig. 9) exhibits a residual strain and a maximum reachable strain that are both fairly well converged within a rather small number of cycles as compared to the situation for the other temperatures. Since the loading at these high temperatures is free of any martensitic transformation, the corresponding cyclic loading involves only the deformation of a non-transforming crystalline phase and of the amorphous phase. This result therefore clearly demonstrates that the transformation and deformation of the crystalline SMA region at the temperatures which do show a martensitic transformation is an indispensable factor for the accumulation of the residual strain with increasing number of cycles, in addition to the plastic deformation of the amorphous pillar parts. As explained above, the plastic deformation of the amorphous region occurring during the first cycle can facilitate the forward transformation and inhibit the reverse transformation of the crystalline SMA region during the second cycle. In the second cycle, the amorphous region can undergo a more severe deformation because the forward transformation of the nearby crystalline region is promoted and thus the maximum reachable strain is increased. In the third cycle, the increased plastic deformation in the amorphous shell can facilitate the forward transformation again. As these processes are repeated, the residual strain gradually accumulates, and the maximum reachable strain gradually increases with increasing loading cycles.

### 3.4 Impact of geometry, maximum stress, and number of cycles

We have shown that the present MD simulations on the compression of pillars including an amorphous region can help to understand experimental results reported for SMAs at the small scale. However, typical experimental conditions are different from the ones achievable in simulations, in particular regarding the absolute sizes of the pillars. In this section, we estimate the influence of the computational limitations by focusing on the cyclic deformation of single crystal pillars with an amorphous shell. Specifically, we analyze stress-strain responses of pillars with different thicknesses of the amorphous shell (0.5 – 3.0 nm) for up to 3 cyclic loadings under different maximum stresses (0.64 – 1.27 GPa). The ratio ($t/d$) between the thickness of the shell ($t$: 2.0, 2.5 and 3.0 nm) and the diameter of the simulated pillars ($d$: 24, 25, and 26 nm) amounts to 0.08, 0.10, and 0.12, respectively.

Figure 12 shows the resulting stress-strain responses. The pillars with relatively thick shells (2.0, 2.5, and 3.0 nm) exhibit stress-strain responses in good consistency with those reported in the previous experiments on NiTi SMA pillars with diameters of 149 – 357 nm [10]. This means concretely that the strain recovery is strongly decreased and that the upper and lower plateaus are almost diminished. We can roughly estimate the $t/d$ ratio of the experimental pillars to be in the range of 0.04 – 0.13, by utilizing a thickness of the amorphous region of $t = 15 – 20$ nm [9, 10]. Thus, the estimated $t/d$ range (0.04 – 0.13) overlaps well with our simulated $t/d$ range (0.08, 0.10, and 0.12) and this overlap explains the similarities in the observed transformation behavior.

Our simulations on the pillars with thinner shell thicknesses (0.5 and 1.0 nm) under relatively large maximum stresses (1.11 and 1.27 GPa) reveal the already detailed sudden stabilization of the martensite phase. As also discussed above, this finding is qualitatively similar to the results of the experiments on CuAlNi SMA pillars with diameters of 550 and 1700 nm [15]. However, even considering the smallest simulated shell thickness ($t$: 0.5 nm), the $t/d$ ratio in the present calculations is 0.02 and therefore about a factor of two larger than the $t/d$ ratio of the experimental submicron- or micron-sized pillars (0.009 – 0.012) [15]. Thus, in order to explain the similarity between the experimental and simulation results despite the difference in the $t/d$ ratio, the difference in the number of cycles needs to be considered. Since the plastic deformation is gradually accumulating during cyclic loading, the number of cycles is an important factor for the sudden occurrence of the retained martensite. The present simulations were performed with a larger $t/d$ ratio, but the maximum number of cycles is only 3 – 5 whereas the experimental value was 100 – 200 [15]. We expect that if the loading is applied to a pillar with a reduced $t/d$ ratio (e.g., similar to the experimental value), more loading cycles will be required to observe the sudden stabilization of the martensite phase as seen in the experiments.

The obtained characteristics of each cycle shown in Fig. 12, e.g., the presence of the phase transformation, presence of retained martensite, and the amount of the residual strain, are summarized in Fig. 13. The presence of the phase transformation and of the retained martensite during each cycle were confirmed by analyzing corresponding atomic configurations. Figure 13(a) indicates that the phase transformation is fully suppressed under conditions when the maximum loading is too low and the proportion of the amorphous region is too large. However, even if the phase transformation does not appear in the first cycle, the phase transformation can appear in later cycles due to the accumulation of plastic deformation as, e.g., for the shell thickness of 1.5 nm and the maximum loading of 0.64 GPa shown in Figs. 12 and 13(a). As shown in Fig. 13(b), retained martensite occurs more readily with increasing cycles and stresses. The amount of accumulated residual strain [Fig. 13(c)] exhibits a similar tendency because the residual strain is greatly influenced by the presence of the retained martensite. The occurrence of the retained martensite and the amount of the accumulated residual strain are most pronounced in pillars with a medium shell thickness (1.5 nm). It seems that the synergetic contribution by the amorphous shell and crystalline SMA regions can be maximized at a certain $t/d$ ratio. This $t/d$ ratio amounts to about 0.065 at a shell thickness of 1.5 nm and diameter of 23 nm. This feature is consistent with reported experimental trends on the cyclic deformation of NiTi nanocrystalline wires by Delville *et al.* [17, 18]. These works [17, 18] investigated the impact of grain size on the cyclic response with direct TEM observations of the dislocation slip for each microstructure obtained after cyclic loading. These works clearly revealed that nanocrystalline wires with finer microstructures (grain diameter < 100 nm) are resistant against dislocation slip, and the amount of accumulated residual strain under cyclic loading is maximized in the wire with a medium grain size among wires with finer microstructures.

**3.5 Methods to mitigate the degradation of superelasticity**

We have revealed that a synergetic interplay of the amorphous and crystalline regions in nano-sized SMAs can cause the experimentally observed degradation of superelasticity under cyclic loading. In this section, based on the gained knowledge, we evaluate previously reported methods and propose a new method to mitigate the degradation of superelasticity. We focus on the cyclic deformation of the single crystal pillar with an amorphous shell.

We first validate two methods of restoring superelasticity reported in a previous experimental study [15] and uncover their working mechanisms. The following two methods were proposed in Ref. [15]: (i) heating to higher temperatures and (ii) additional mechanical loading under a reduced maximum stress. Figure 14 shows the verification of method (i) by MD simulations of mechanical and thermal cyclic loading using a pillar with a diameter of 20 nm and shell thickness of 0.5 nm. The initial mechanical cycles ($1^{st} - 3^{rd}$) of compressive loading and unloading at 450 K exhibit the already mentioned sudden increase in the residual strain [blue lines in Fig. 14(a)] due to the formation of retained martensite, as explicitly confirmed by the atomic configurations in Fig. 14(c). In the intermediate thermal cycles ($4^{th} - 6^{th}$) of heating to 650 K and cooling to 450 K at zero stress, the residual strain is substantially reduced [Fig. 14(b)] because the retained martensite is successfully eliminated [Fig. 14(c)]. In the final mechanical cycles ($7^{th} - 8^{th}$) the superelastic response has been recovered showing again stress plateaus due to the transformation and a decreased residual strain. These results confirm that thermal annealing after mechanical loading is indeed effective to mitigate the degradation of superelasticity. In fact, the previous experiments [15] focused on a single heating and cooling step, but our results indicate that multiple heating and cooling cycles can be even more effective. As shown in Fig. 14(b) the accumulated residual strain is mostly resolved in the first thermal cycle, but subsequent thermal cycles contribute cumulatively. We thus propose an improvement of method (i), i.e., an increase of the number of restoring thermal cycles.

We have also examined method (ii), i.e., the reported peculiar behavior that superelasticity can be recovered by additional mechanical loading albeit with a reduced maximum stress [15]. Figure 15 shows the stress-strain response of a pillar with a diameter of 20 nm and a shell thickness of 0.5 nm under stepwise mechanical loading. The initial three cycles ($1^{st} - 3^{rd}$) were performed with a maximum stress of 1.27 GPa at 450 K. They are thus exactly the same as the initial cycles in Fig. 14 and likewise indicate a sudden increase in the residual strain [blue lines in Figs. 15(a-c)] due to the formation of retained martensite [Fig. 15(d)]. The subsequent 10 mechanical cycles ($4^{th} – 13^{th}$) were performed at significantly lower maximum stress at the same temperature. Specifically, we have tested stresses of 0.48 GPa, 0.38 GPa and 0.29 GPa [Fig. 15(a-c)]. The residual strain accumulated during the initial 3 cycles can be gradually relieved during the subsequent cycles. The rate of recovery of the superelastic response depends on the maximum stress of the subsequent cycles. The cyclic loading with a maximum stress of 0.29 GPa [Fig. 15(c)] reduces the residual strain more easily as compared to that of 0.48 GPa [Fig. 15(a)]. However, the residual strain is not completely removed and appears to converge to the amount of residual strain that occurred in the first cycle of the loading. The gradual recovery of superelasticity during the lower-stress cycles is an opposite behavior to the negative synergetic degradation process of the former high-stress cycles. When cyclic loading at a lower maximum stress is conducted to a pillar with retained martensite, the maximum reachable strain of a cycle decreases, and this causes a reduced deformation in the amorphous shell region. A decrease in the deformation of the amorphous region causes a decrease in the maximum reachable strain again in the next cycle. As these processes are repeated, the residual strain steadily decreases, and the retained martensite disappears with increasing cycles as shown in Fig. 15(d).

As another possible method to mitigate the degradation of superelasticity of small-scale SMAs, one can think of a pre-training before their final use. In bulk SMAs, pre-training by cyclic loading is known to reduce the unwanted residual strain and to stabilize the hysteretic response in practical applications [3, 4]. A difficulty for the present case of SMAs containing an amorphous region at the small scale is that the usual pre-training under cyclic loading itself can activate the synergetic degradation process that accumulates the residual strain. Based on the mechanism identified in the present study, we have been able to circumvent this difficulty and to devise a simple but effective pre-training method of SMAs to reduce the degradation of the superelasticity during practical operation. Since the degradation of the superelasticity is caused by the synergetic contribution of the plastic deformation of the amorphous region and the deformation due to variances in twin structure of the martensite phase in the crystalline SMA region, the crucial point is to break this synergetic process during cyclic loading. The principle of the new training method is to perform the training such as to avoid the phase transformation of the crystalline region by selecting a higher temperature than the actual operating temperature of small sized SMAs.

We have validated our new pre-training method by performing additional MD simulations of mechanical cyclic loading using a pillar with a diameter of 20 nm and a shell thickness of 0.5 nm as shown in Fig. 16. When the initial cyclic loading ($1^{st} – 3^{rd}$) for the pre-training [orange lines in Fig. 16 (a)] is conducted at 650 K, the stress-strain response shows no sign of the martensitic transformation, as also confirmed by the atomic configurations in Fig. 16(b). During these cycles ($1^{st} – 3^{rd}$), the residual strain and the maximum reachable strain are well converged without inducing the synergetic degradation associated with the phase transformation and the deformation of the amorphous shell. If subsequent cyclic loading ($4^{th} - 8^{th}$) is conducted at a decreased temperature of 450 K [blue lines in Fig. 16(a)], the stress-strain response exhibits a stable superelastic response over the entire cycle with a decreased residual strain and without the presence of any retained martensite as clarified in Fig. 16(b). The effectiveness of the pre-training is impressively shown by comparing the stress-strain responses with the pre-training [blue lines in Fig. 16(a)] and that without the pre-training [blue lines in Figs. 14(a) and 15(a-c)]. These results demonstrate that the devised pre-training method is very effective in maintaining the cyclic stability of small-scale SMAs.

## 4. Conclusions

Large-scale MD simulations have been performed to provide an atomic-scale understanding of the degradation of superelasticity of SMAs at the small scale under cyclic compression. Our work reveals that a possible factor responsible for the observed degradation of the superelasticity under cyclic loading is the accumulated plastic deformation due to a synergetic interplay of the amorphous and crystalline SMA regions. Based on the identified mechanism, we have validated previously reported methods of recovering the superelasticity and have proposed a novel design strategy for achieving a sustainable operation of SMAs at the small scale. Our results can be summarized as follows:

(1) The stress-strain response of pristine SMA pillars indicates an almost complete hysteresis loop without any noticeable residual strain. Similar to bulk SMAs, overall stresses increase with increasing temperature, and the phase transformation is totally inhibited for too high temperatures.

(2) The applied preparation procedure to introduce an amorphous region in the pillars is crucial to achieve stress-strain responses with features as observed in previous experiments. The increase in the residual strain with decreasing pillar size and the sudden loss of superelasticity under cyclic loading are faithfully reproduced by the present MD simulations.

(3) An amorphous shell has a qualitatively similar impact on transformation stresses and temperatures as an amorphous-like grain boundary region, but the latter shows a stronger quantitative impact, e.g., the transformation temperatures decrease by about 100 K more than for the pillar with the amorphous shell.

(4) The two main sources of the accumulated residual strain are the plastic deformation of the amorphous region and the deformation due to variances in twin structure of the martensite phase in the crystalline SMA region. During cyclic loading, both effects contribute synergistically to the accumulation of the residual strain.

(5) The amount of residual strain under cyclic loading is maximized at a certain ratio between the amorphous and crystalline SMA regions which maximizes the synergetic degradation effect. When the relative thickness of the amorphous shell is too large, the contribution of the amorphous region to the residual strain is large, but the contribution of the SMA region is limited due to the suppression of the phase transformation.

(6) The sudden loss of the superelasticity of pillars under cyclic loading can be partially resolved by heating to a higher temperature or by applying subsequent cyclic loading at a lower maximum stress. We have shown that these methods cause a healing process which is inverse to the synergetic degradation process during the initial cycles, resulting in a reverse transformation of the retained martensite.

(7) We have proposed and successfully validated a new method to mitigate the degradation of the superelasticity. The present MD results suggest that a pre-training of small sized SMAs at higher temperatures avoids the phase transformation and is thus effective to break the negative impact of the synergetic contribution by the amorphous shell and crystalline SMA regions.

**Appendix**

Figure 17 shows the benchmark simulations to evaluate the impact of the aspect ratio (height/diameter) and the loading rate on the stress-strain response. Figure 18 shows the temperature dependence of the engineering strain of SMA pillars under cooling and reheating, resulting from different preparation conditions (different diameters and thicknesses of the amorphous region). MD simulations for the temperature-induced phase transformation were performed in the isobaric-isothermal ($NPT$) ensemble at zero pressure. Starting at 500 K, the temperature was gradually decreased to 20 K and increased again to 500 K with cooling and heating rates of ±0.5 K/ps.

**Acknowledgements**

The funding by the European Union's Horizon 2020 research and innovation programme (Grant Agreement No. 639211) is gratefully acknowledged. This work was carried out under the framework of the international cooperation program managed by the National Research Foundation of Korea (Grant No. NRF-2019K2A9A2A06020258) and National Natural Science Foundation of China (Grant No. 51911540474). This work was also supported by the National Research Foundation of Korea (NRF) funded by the Ministry of Science and ICT (Grant No. NRF-2019R1F1A1040393, NRF-2019M3D1A1079214 and NRF-2019M3E6A1103984), and the National Supercomputing Center with supercomputing resources including technical support (KSC-2019-CRE-0042).

Table 1 Dimensions of the simulation cells and corresponding numbers of atoms considered in the present MD simulations.

| | Pristine pillar | | | | Pillar with amorphous shell | | | | | | | | | Nano-crystalline pillar |
|---|---|---|---|---|---|---|---|---|---|---|---|---|---|---|
| Thickness of shell (nm) | 0.0 | 0.0 | 0.0 | 0.0 | 0.5 | 0.5 | 0.5 | 0.5 | 1.0 | 1.5 | 2.0 | 2.5 | 3.0 | 0.0 |
| Diameter of pillar (nm) | 15 | 20 | 25 | 30 | 16 | 21 | 26 | 31 | 22 | 23 | 24 | 25 | 26 | 30 |
| Height of pillar (nm) | 15 | 20 | 25 | 30 | 15 | 20 | 25 | 30 | 20 | 20 | 20 | 20 | 20 | 30 |
| Number of atoms | $1.9\times10^5$ | $4.6\times10^5$ | $9.0\times10^5$ | $1.5\times10^6$ | $2.2\times10^5$ | $5.1\times10^5$ | $9.7\times10^5$ | $1.7\times10^6$ | $5.6\times10^5$ | $6.1\times10^5$ | $6.7\times10^5$ | $7.2\times10^5$ | $7.8\times10^5$ | $1.5\times10^6$ |

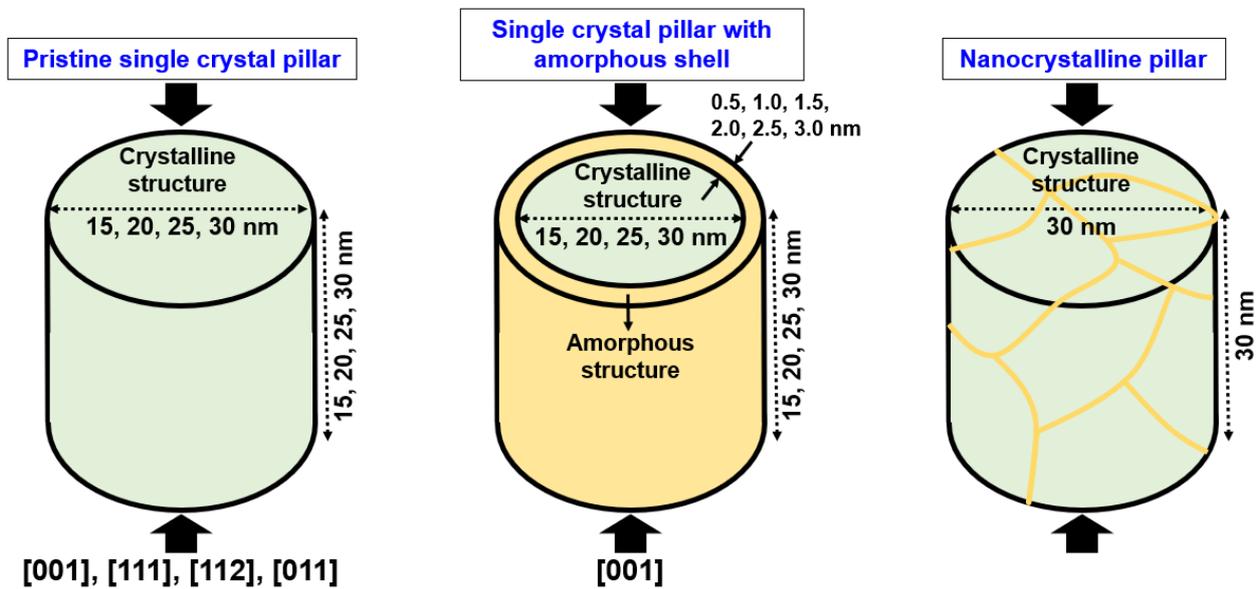

Fig. 1 Schematic illustration of the NiTi SMA pillars used in the present MD simulations. The aspect ratio between the height ($h$) of a pillar and the diameter ($D$) of the crystalline structure ($h/D$) is set to 1. A nanocrystalline pillar was generated by cutting an initial cube-shaped bulk nanocrystalline cell with 5 grains.

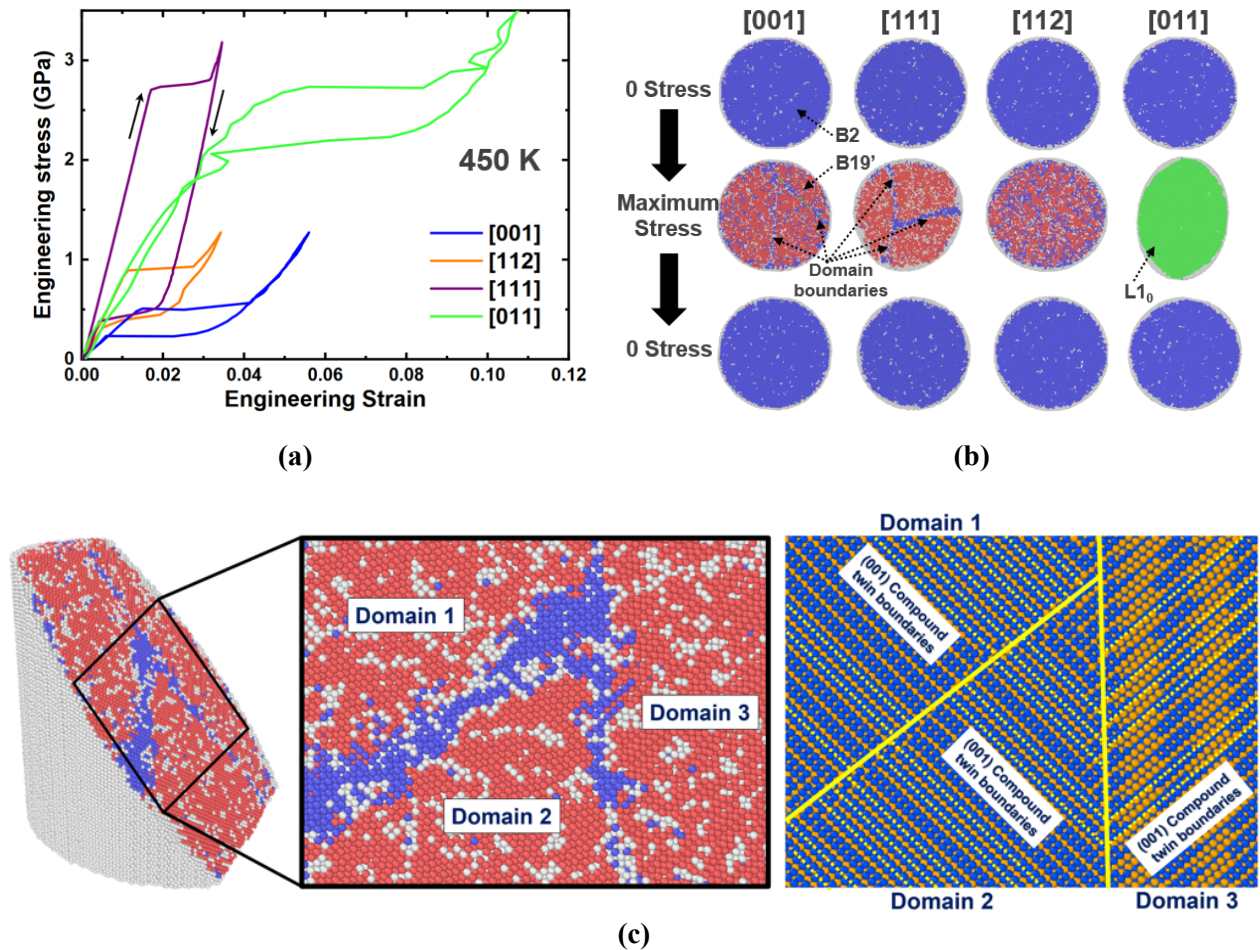

Fig. 2 (a) Stress-strain responses of pristine single crystal pillars with a diameter of 20 nm under compressive loading and unloading along the [001], [112], [111] and [011] directions at 450 K. (b) Corresponding atomic configurations of pillars at the initial state (zero stress), at the maximum compressive loading, and after the unloading process (zero stress). The color of the atoms is scaled according to the AC-CNA pattern. In each configuration, blue atoms correspond to the B2 austenite structure, red atoms to the B19' martensite structure, light green atoms to the $L1_0$ structure, and gray atoms to the surface region and amorphous structure. (c) Visualization of the domain and twin structures of the pillar after the martensitic transformation loaded along the [111] direction. The subfigure on the left is visualized by the AC-CNA pattern. In the subfigure on the right, Ni and Ti atoms are represented by blue and orange colors, respectively.

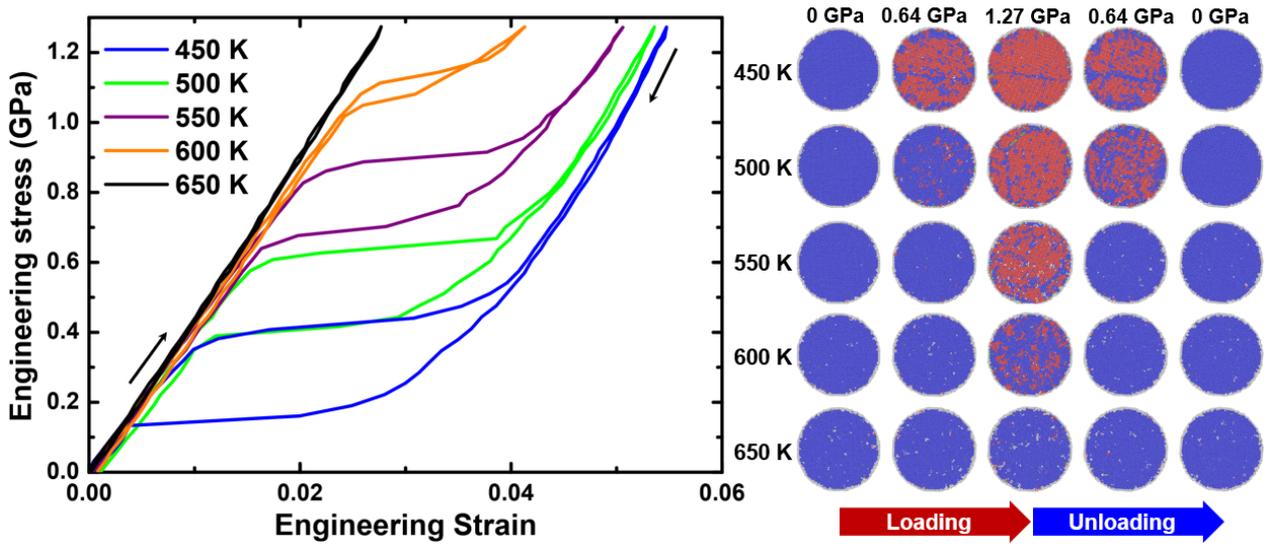

Fig. 3  Stress-strain response of a pristine single crystal pillar ([001] orientation) with a diameter of 15 nm under compressive loading and unloading at different temperatures. Corresponding atomic configurations of the pillar at different levels of the stress are shown to the right. The color of the atoms is scaled according to the PTM pattern, where blue atoms correspond to the B2 austenite structure, red atoms to the B19' martensite structure, and gray atoms to the surface region and amorphous structure.

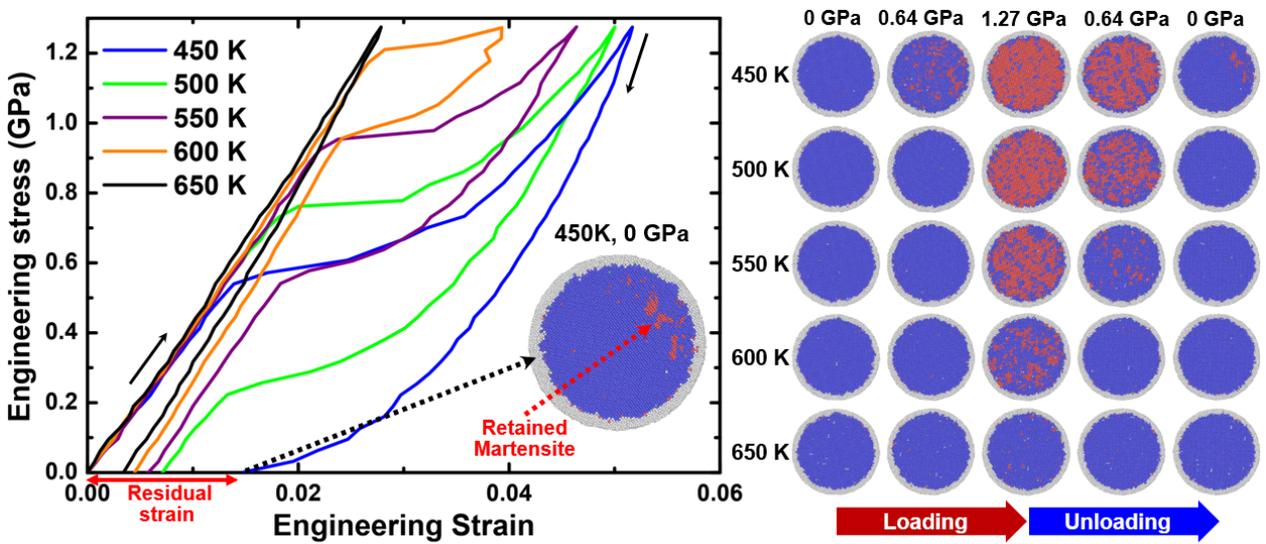

Fig. 4  Stress-strain response of a single crystal pillar ([001] orientation; diameter of the crystalline region: 15 nm) with an amorphous surface shell (thickness: 0.5 nm) under compressive loading and unloading at different temperatures. Corresponding atomic configurations of the pillar at different levels of the stress are visualized by the PTM pattern with the same color coding as in Fig. 3.

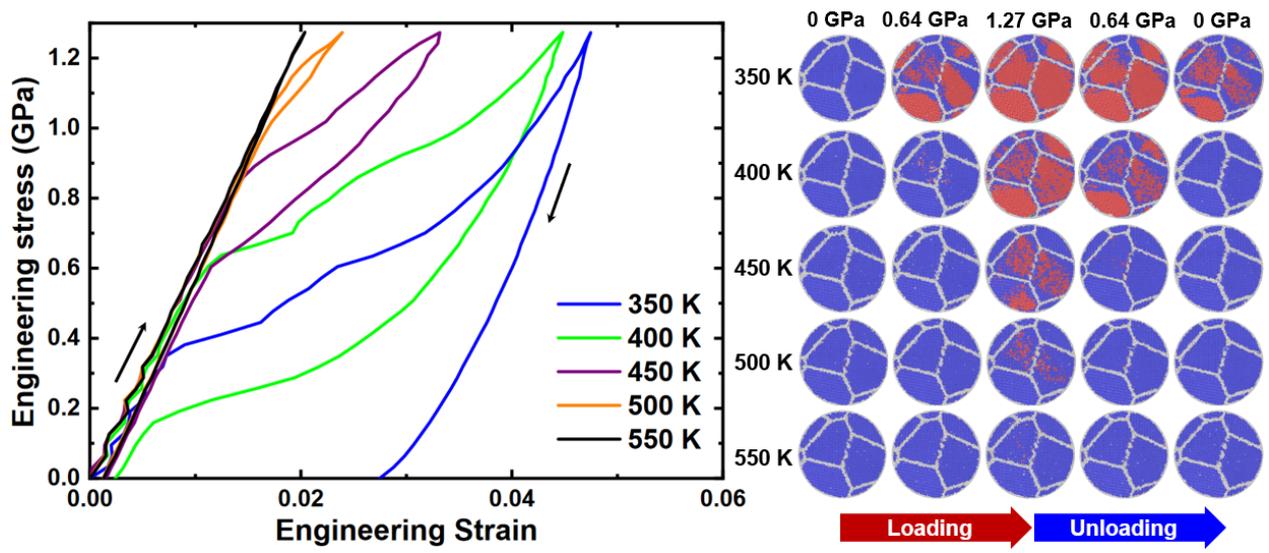

Fig. 5  Stress-strain response of a nanocrystalline pillar with a diameter of 30 nm under compressive loading and unloading at different temperatures. Corresponding atomic configurations of the pillar at different levels of the stress are visualized by the PTM pattern with the same color coding as in Fig. 3.

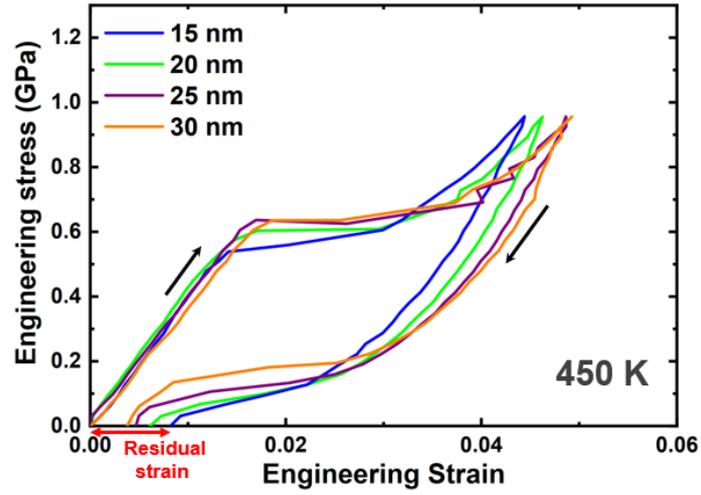

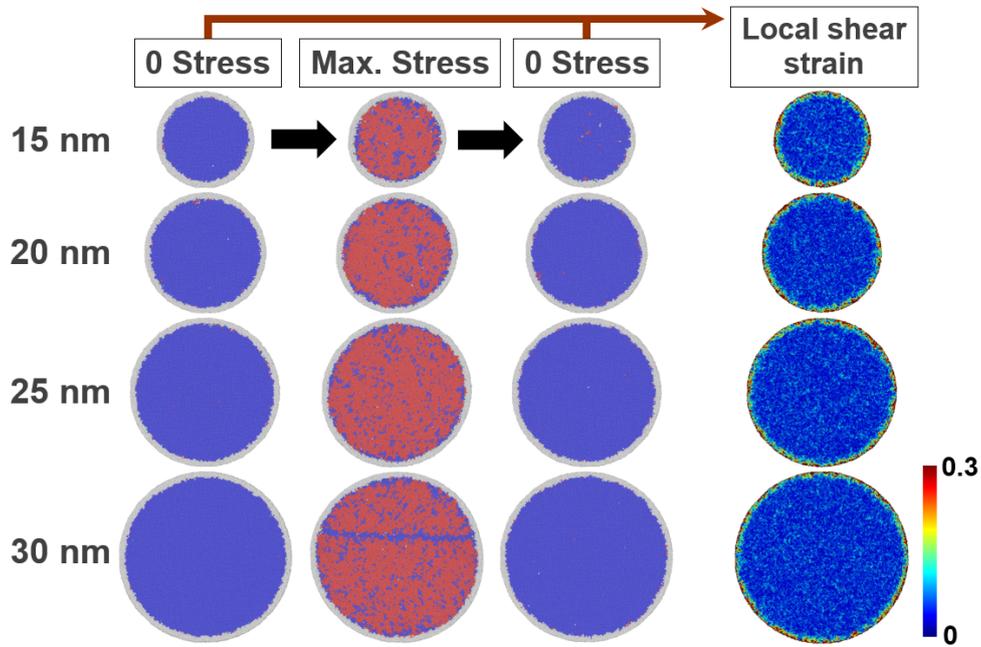

Fig. 6  (a) Stress-strain responses of single crystal pillars with varying diameter of the crystalline region (15, 20, 25 and 30 nm) with an amorphous surface shell (thickness: 0.5 nm) under compressive loading and unloading at 450 K. (b) Corresponding atomic configurations at the initial state (zero stress), at the maximum compressive loading, and after the unloading process (zero stress), visualized by the PTM pattern with the same color coding as in Fig. 3. In the most right atomic snapshots, the color of the atoms is scaled according to the local shear strain [51].

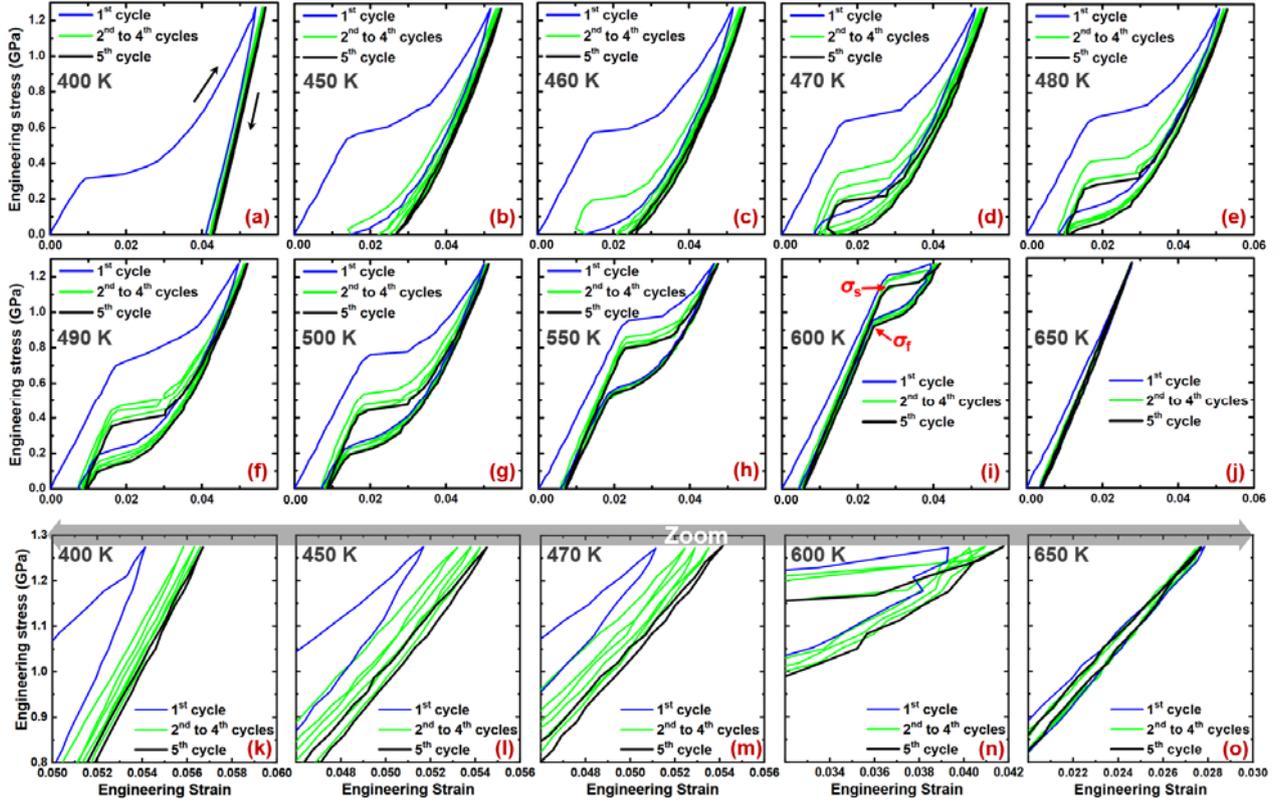

Fig. 7 (a-j) Stress-strain response of a single crystal pillar ([001] orientation; diameter of the crystalline region: 15 nm) with an amorphous surface shell (thickness: 0.5 nm) under 5 cyclic compressive loadings at different temperatures. Examples of the critical stress at the start of the forward transformation ($\sigma_s$) and the critical stress at the end of the back transformation ($\sigma_f$) are indicated by the red arrows in Fig. (i). (k-o) Magnified figures of (a), (b), (d), (i) and (j) highlighting the variation of the maximum reachable strain during cyclic loading.

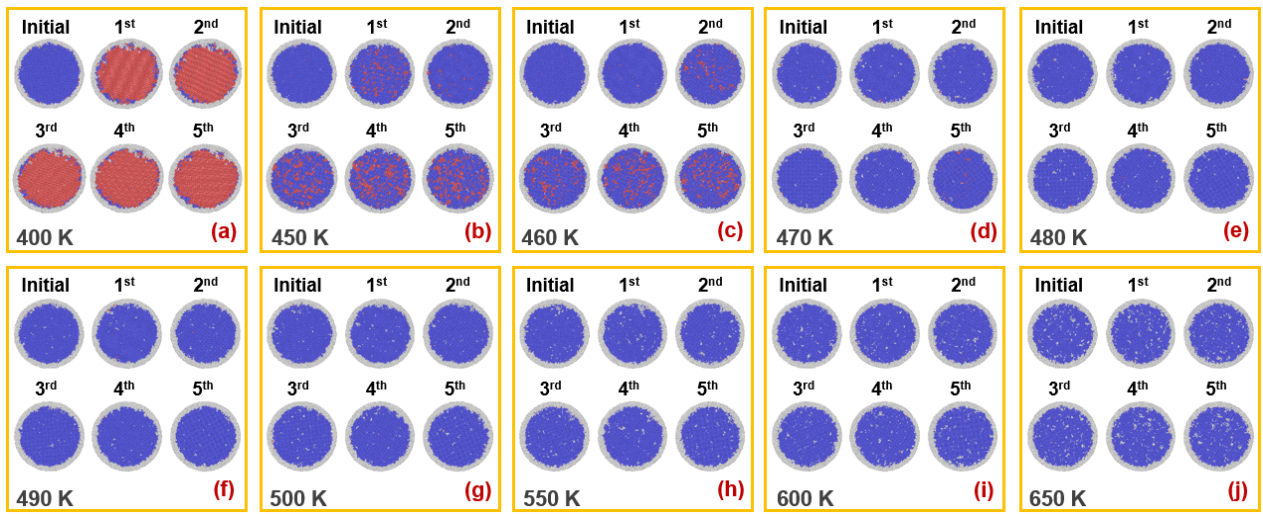

Fig. 8 Atomic configurations corresponding to the various stress-strain responses of the single crystal pillar shown in Figs. 7(a-j). Configurations after each loading cycle (zero stress) are visualized by the PTM pattern with the same color coding as in Fig. 3.

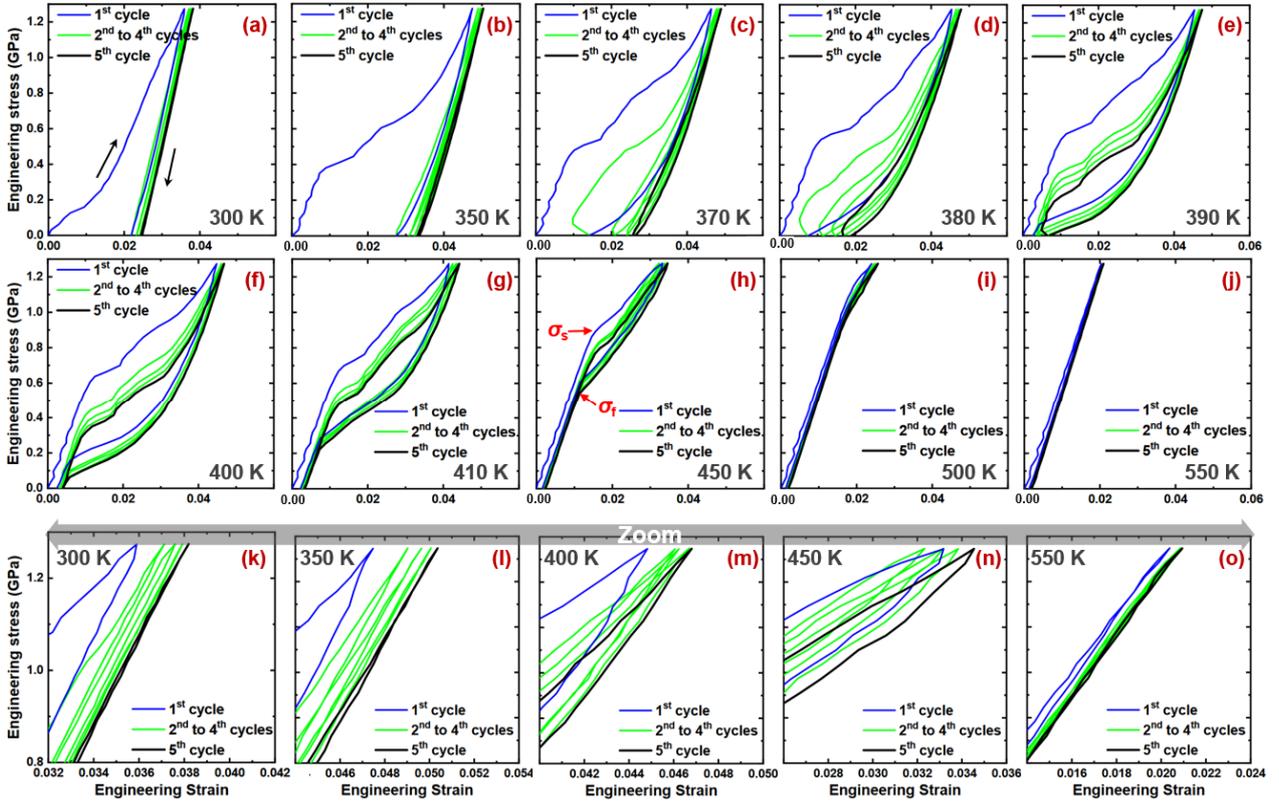

Fig. 9 (a-j) Stress-strain response of a nanocrystalline pillar with a diameter of 30 nm under 5 cyclic compressive loadings at different temperatures. Examples of the critical stress at the start of the forward transformation ($\sigma_s$) and the critical stress at the end of the back transformation ($\sigma_f$) are indicated by the red arrows in Fig. (i). (k-o) Magnified figures of (a), (b), (d), (i) and (j) highlighting the variation of the maximum reachable strain during cyclic loading.

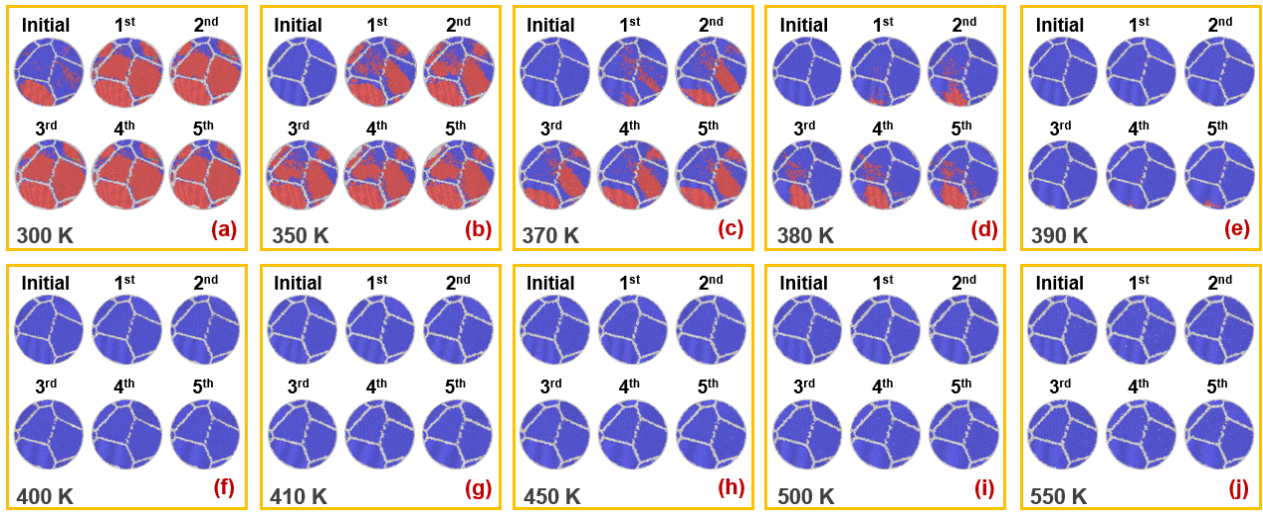

Fig. 10 Atomic configurations corresponding to the various stress-strain responses of the nanocrystalline pillar shown in Figs. 9(a-j). Configurations after each loading cycle (zero stress) are visualized by the PTM pattern with the same color coding as in Fig. 3.

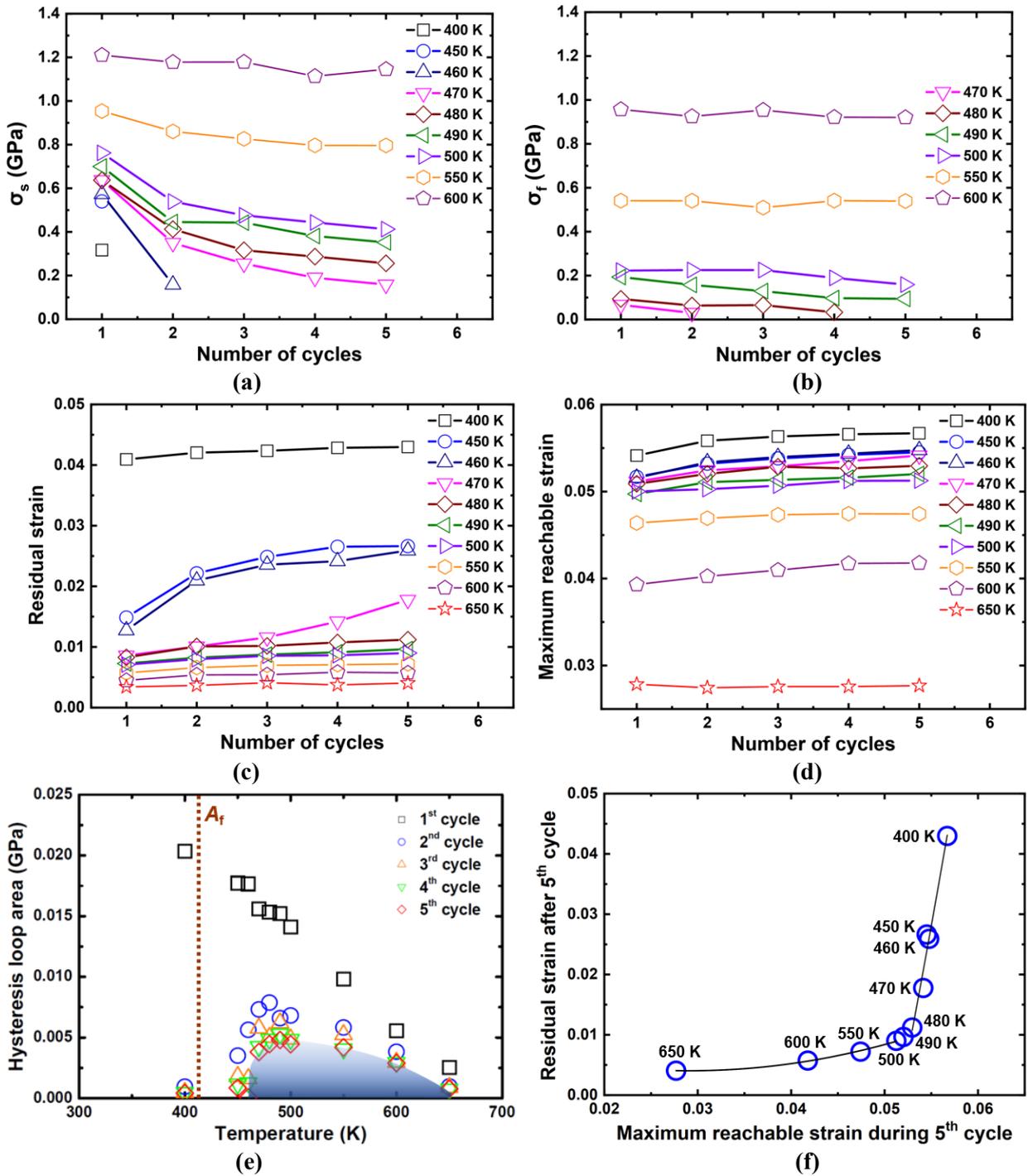

Fig. 11 Summary of the key properties extracted from the stress-strain behavior of the single crystal pillar shown in Fig. 7. (a) Critical stresses at the start of the forward transformation ($\sigma_s$), (b) critical stresses at the end of the reverse transformation ($\sigma_f$), (c) accumulated residual strains, and (d) maximum reachable strains as a function of the number of cycles. (e) Temperature dependence of the hysteresis loop area calculated for each cycle. The shaded area indicates the condition where the hysteresis loop is stable during 5 cyclic loadings. (f) Correlation between the residual strain after the 5$^{th}$ cycle and the maximum reachable strain (the maximum strain at the maximum stress) during the 5$^{th}$ cycle. Lines connecting symbols in Figs. (a) and (b) are presented only as a guide for the eye. Lines in Fig. (f) are results of regression using a quadratic polynomial function.

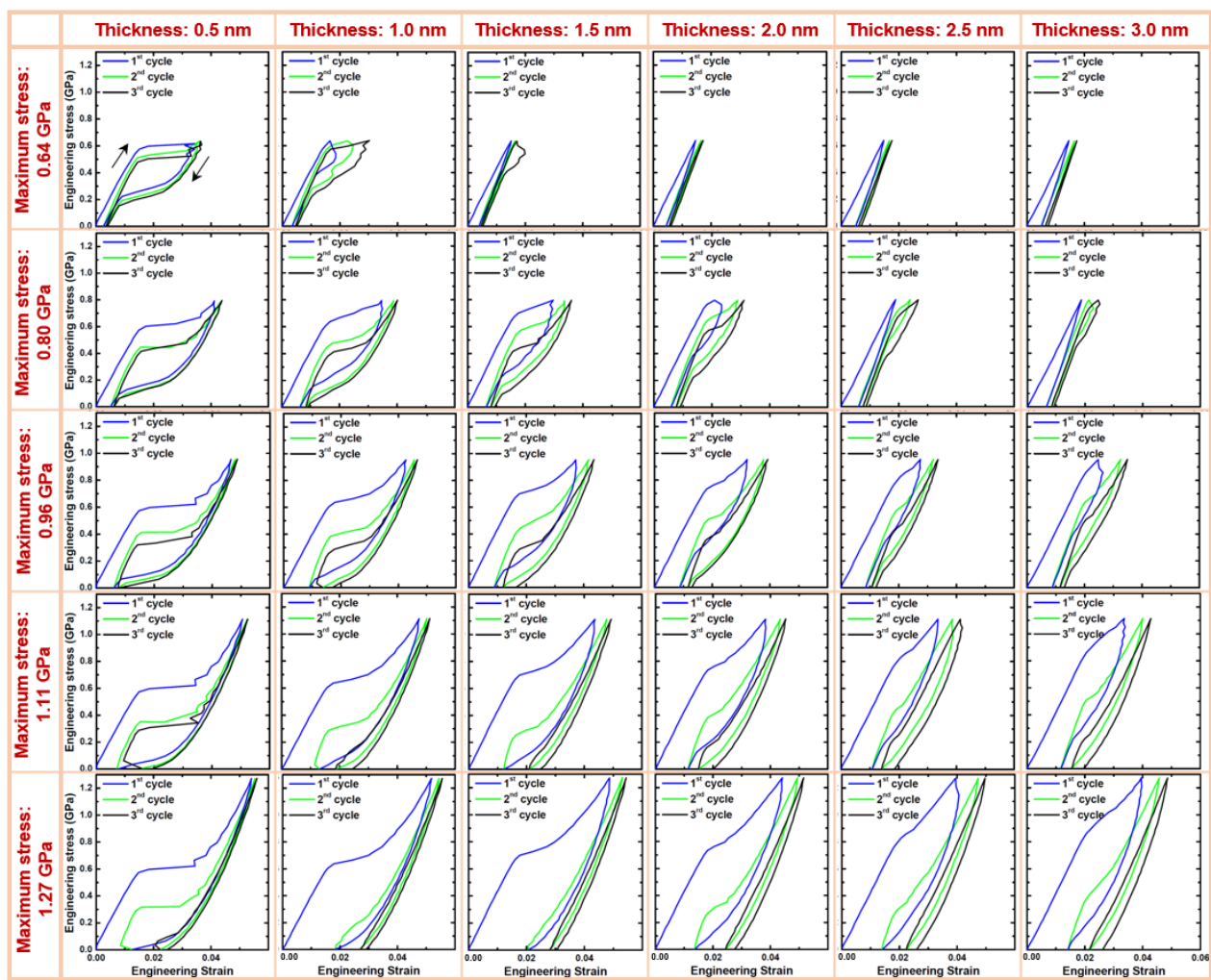

Fig. 12 (a) Stress-strain responses of single crystal pillars (diameter of the crystalline region: 20 nm) with an amorphous surface shell (thickness: 0.5, 1.0, 1.5, 2.0, 2.5 and 3.0 nm) under 3 cyclic loadings (maximum stress: 0.64, 0.80, 0.96, 1.11 and 1.27 GPa) at 450 K.

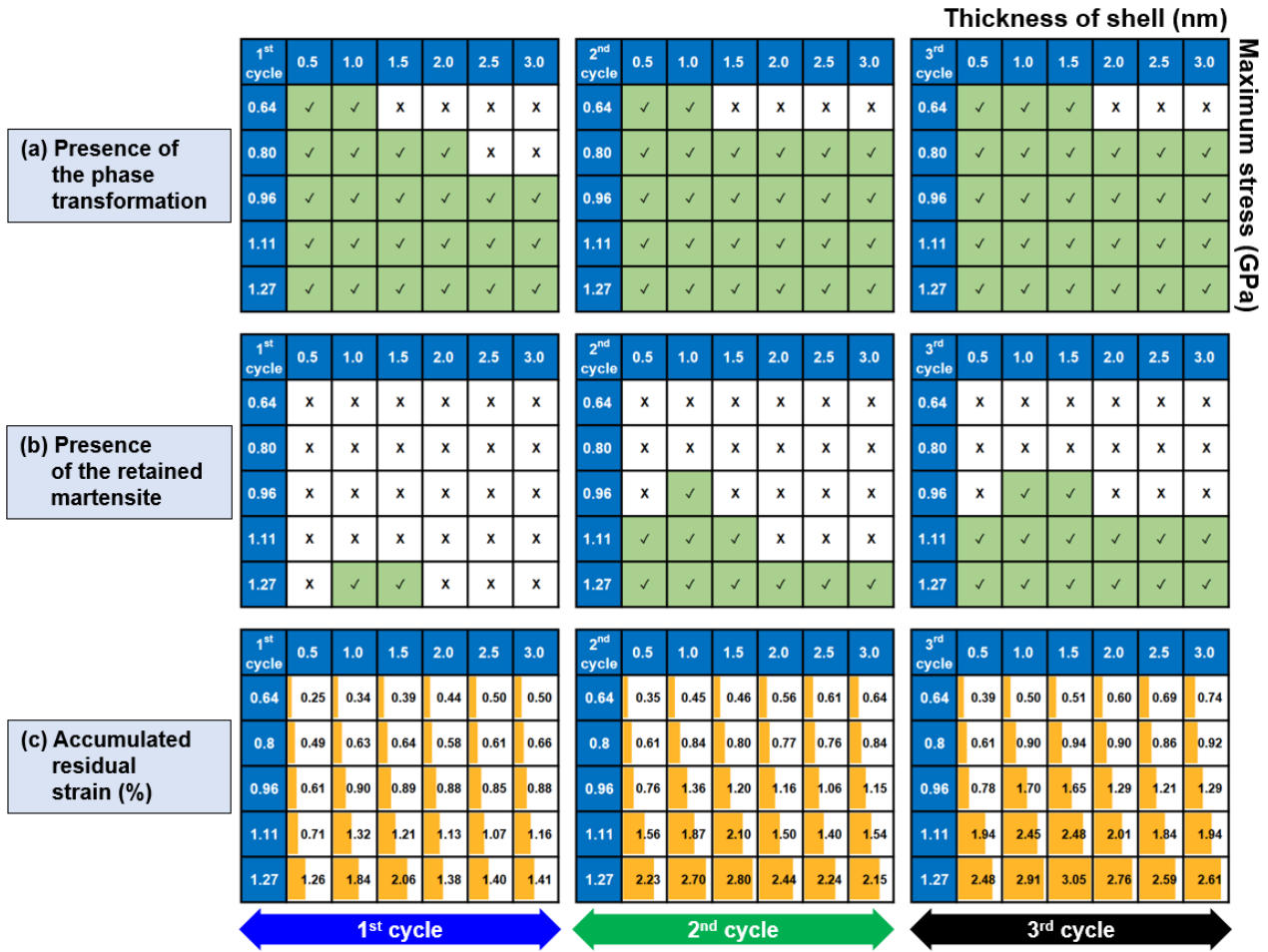

Fig. 13 Summary of the properties obtained from the stress-strain responses of the single crystal pillars shown in Fig. 12. (a) The presence of the martensitic phase transformation during 3 cyclic loadings. (b) The presence of retained martensite at the unloading state (zero stress) after each cycle. "✓" and "X" mean the presence and absence, respectively. (c) The amount of the accumulated residual strain at the unloading state (zero stress) after each cycle.

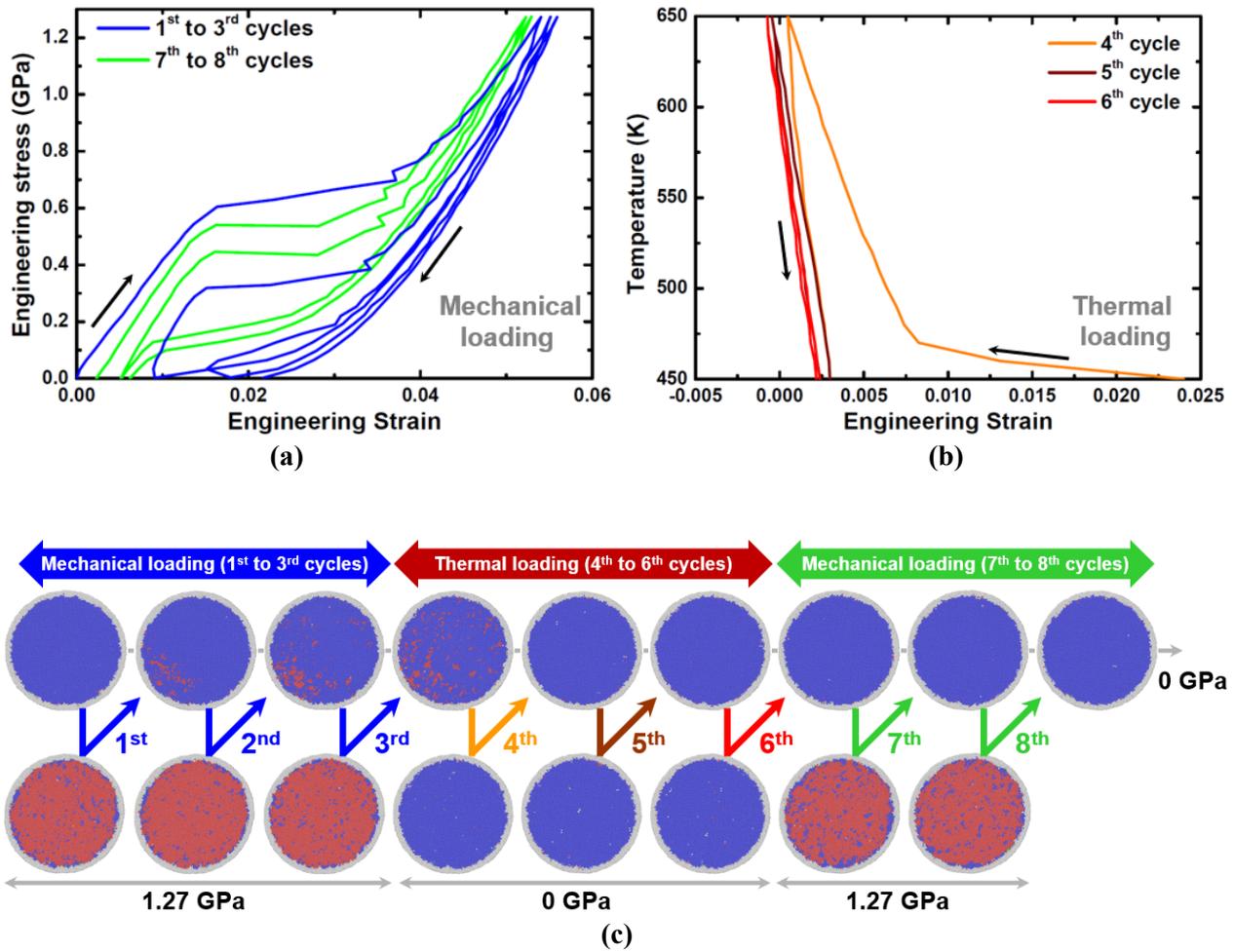

Fig. 14 Stress-strain response of a single crystal pillar (diameter of the crystalline region: 20 nm) with an amorphous surface shell (thickness: 0.5 nm) during (a) initial ($1^{st}$ – $3^{rd}$) and final ($7^{th}$ – $8^{th}$) mechanical cycles. (b) Temperature dependence of the engineering strain during intermediate thermal cycles ($4^{th}$ – $6^{th}$). The mechanical loadings were performed at 450 K with a maximum stress of 1.27 GPa. The thermal loadings were performed under the full relaxation of the pillar (zero stress). (c) Corresponding atomic configurations of the pillar during each cycle are visualized by the PTM pattern with the same color coding used in Fig. 3.

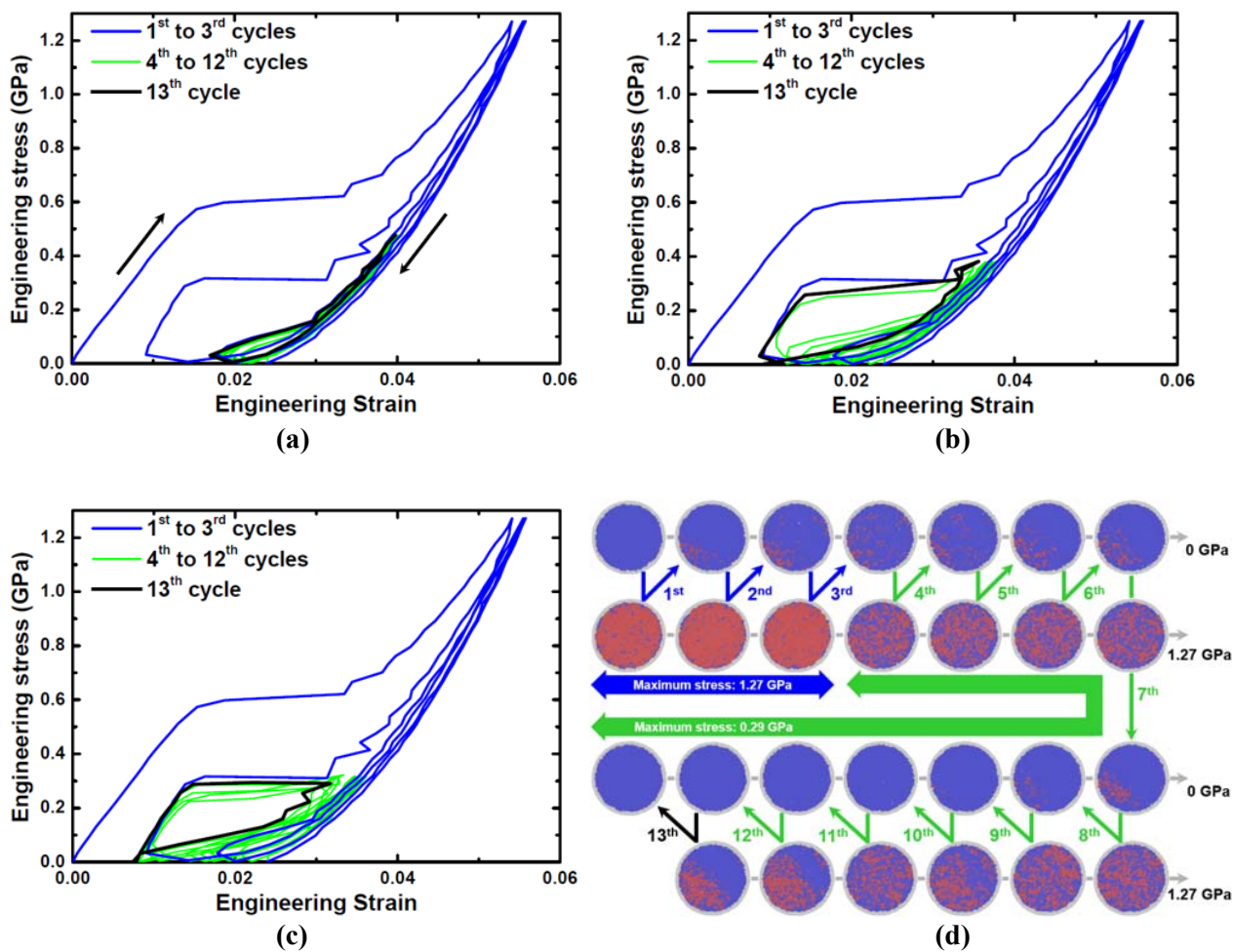

Fig. 15 Stress-strain response of a single crystal pillar (diameter of the crystalline region: 20 nm) with an amorphous surface shell (thickness: 0.5 nm) under stepwise mechanical loadings of initial ($1^{st} - 3^{rd}$) and final ($4^{th} - 13^{th}$) cycles at 450 K. In each response, the maximum stress of initial cycles was commonly set to 1.27 GPa while that of final cycles was set differently to (a) 0.48 GPa, (b) 0.38 GPa, and (c) 0.29 GPa. (d) Atomic configurations of the pillar during each cycle corresponding to (c) are visualized by the PTM pattern with the same color coding used in Fig. 3.

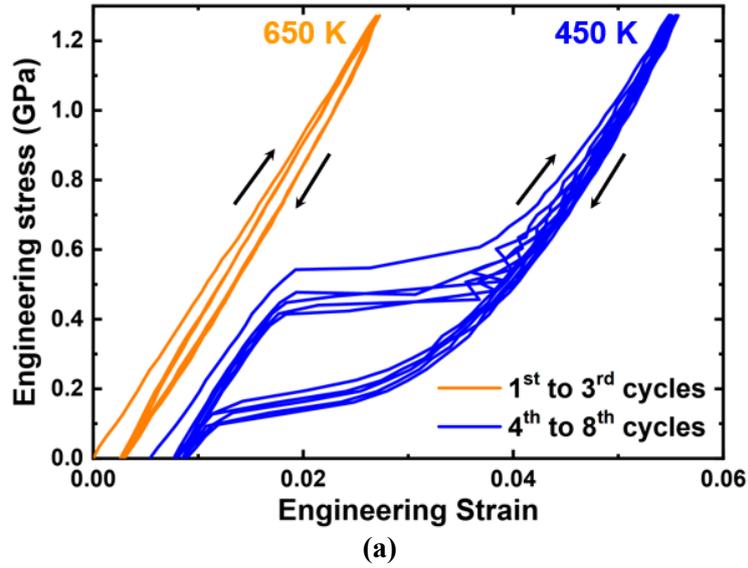

(a)

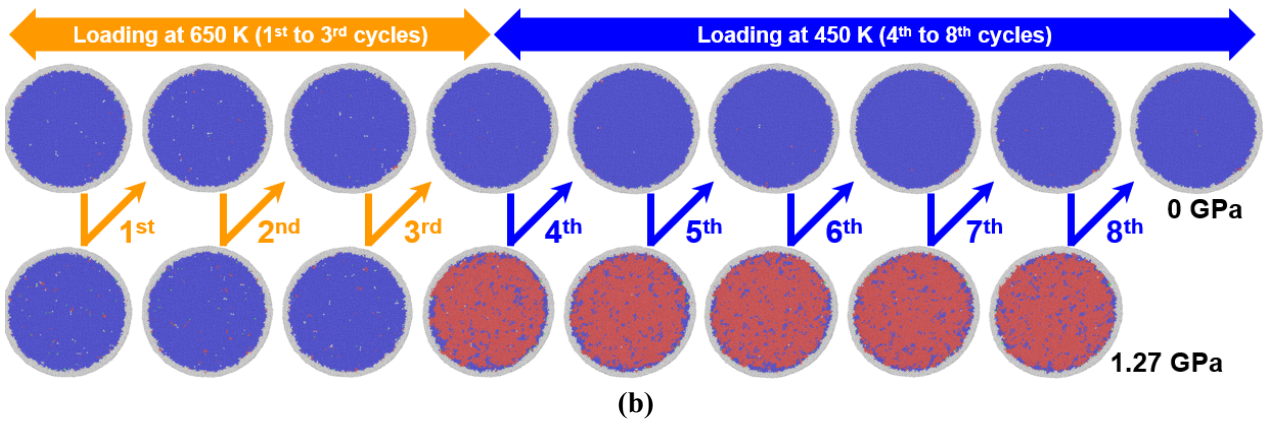

(b)

Fig. 16 (a) Stress-strain response of a single crystal pillar (diameter of the crystalline region: 20 nm) with an amorphous surface shell (thickness: 0.5 nm) under stepwise mechanical loadings of initial ($1^{st}$ – $3^{rd}$) cycles at 650 K and final ($4^{th}$ – $8^{th}$) cycles at 450 K. (b) Corresponding atomic configurations of the pillar during each cycle are visualized by the PTM pattern with the same color coding used in Fig. 3.

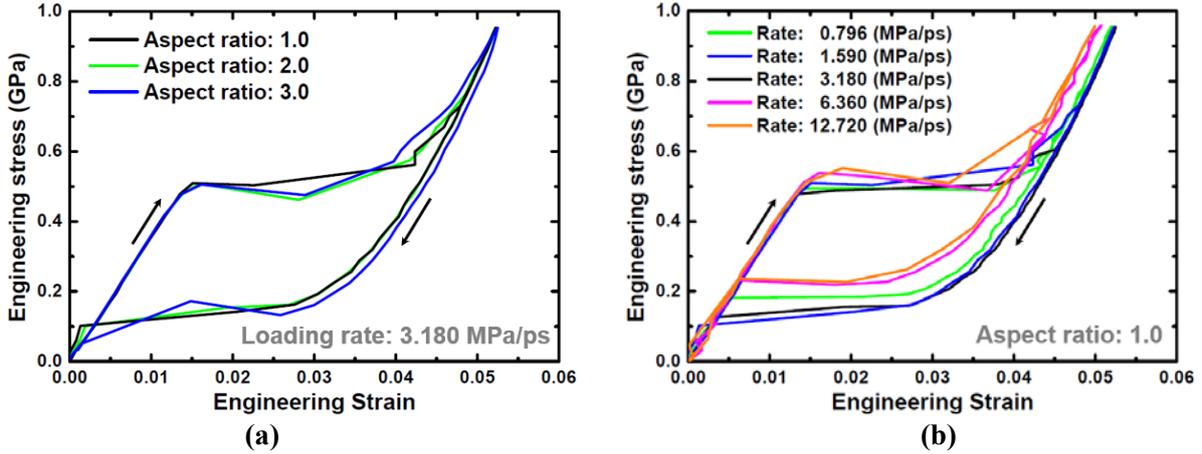

Fig. 17 Benchmark simulations to evaluate effects of (a) the aspect ratio (height/diameter) and (b) the loading rate on the stress-strain responses of pristine single crystal pillars with a diameter of 20 nm at 450 K.

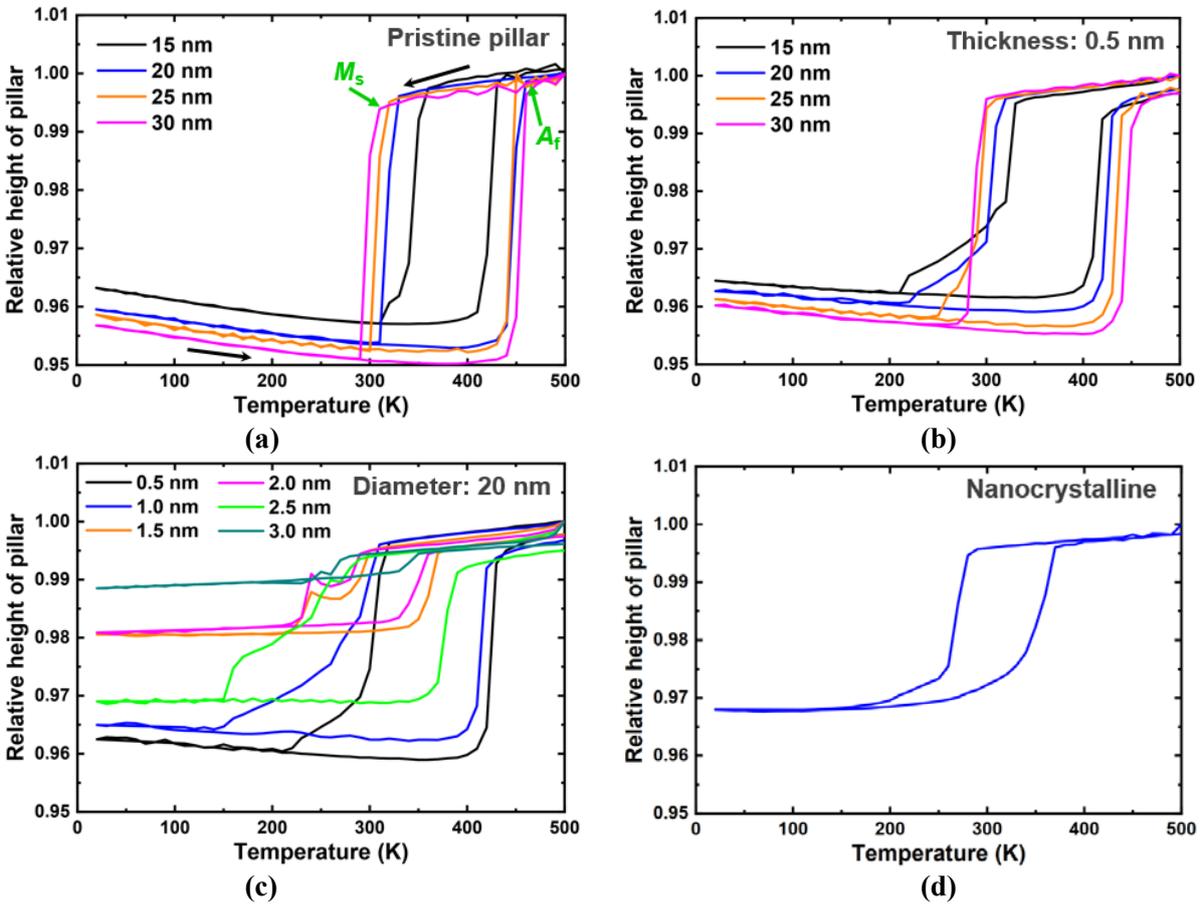

Fig. 18 Temperature dependence of the relative height of pillars under thermal loading. The height of each pillar at 500 K is set as a reference. The discrete jumps represent the occurrence of the martensitic phase transformation. An example martensite start ($M_s$) and the austenite finish ($A_f$) temperature are indicated by the green arrows. (a) Pristine single crystal pillars with different diameters (15, 20, 25 and 30 nm). (b) Single crystal pillars with different diameters (15, 20, 25 and 30 nm) of the crystalline region, in which the amorphous surface shell is added with a thickness of 0.5 nm. (c) Pillars with a diameter of 20 nm, in which the amorphous surface shell is added with different thicknesses (0.5, 1.0, 1.5, 2.0, 2.5 and 3.0 nm). (d) A nanocrystalline pillar with a diameter of 30 nm (5 grains in the initial bulk state) in (a).